\pdfoutput = 1

\documentclass[
reprint,
superscriptaddress,
showkeys,
amsmath,
amssymb,
prb,
floatfix,
]{revtex4-2}

\usepackage{soul}
\usepackage{graphicx}
\usepackage{dcolumn}
\usepackage{bm}
\usepackage{amsthm}
\usepackage{enumitem}
\usepackage{array}
\usepackage{tabularx}
\usepackage{siunitx}
\usepackage[caption=false]{subfig}
\usepackage{placeins}
\usepackage{float}
\usepackage{hyperref}
\graphicspath{{./Figures/}}

\begin{document}

\preprint{APS/123-QED}


\title{Energy landscape of noncollinear exchange coupled magnetic multilayers}

\author{George Lertzman-Lepofsky}
\email{gmlertzm@sfu.ca}\altaffiliation{these authors contributed equally to this work.}
\affiliation{Simon Fraser University, Vancouver, British Columbia V5A 1S6, Canada}

\author{Afan Terko}
\email{ata159@sfu.ca}\altaffiliation{these authors contributed equally to this work.}
\affiliation{Simon Fraser University, Vancouver, British Columbia V5A 1S6, Canada}

\author{Sabri Koraltan}
\affiliation{Faculty of Physics, University of Vienna, Vienna 1010, Austria}

\author{Dieter Suess}
\affiliation{Faculty of Physics, University of Vienna, Vienna 1010, Austria}

\author{Erol Girt}
\email{egirt@sfu.ca}
\altaffiliation{these authors jointly supervised.}
\affiliation{Simon Fraser University, Vancouver, British Columbia V5A 1S6, Canada}

\author{Claas Abert}
\email{claas.abert@univie.ac.at}
\altaffiliation{these authors jointly supervised.}
\affiliation{Faculty of Physics, University of Vienna, Vienna 1010, Austria}


\begin{abstract}

We conduct an exploration of the energy landscape of two coupled ferromagnetic layers with perpendicular-to-plane uniaxial anisotropy using finite-element micromagnetic simulations. These multilayers can be used to produce noncollinearity in spin-transfer torque magnetic random-access memory cells, which has been shown to increase the performance of this class of computer memory. We show that there exists a range of values of the interlayer exchange coupling constants for which the magnetic state of these multilayers can relax into two energy minima. The size of this region is determined by the difference in the magnitude of the layer anisotropies and is minimized when this difference is large. In this case, there is a wide range of experimentally achievable coupling constants that can produce desirable and stable noncollinear alignment. We investigate the energy barriers separating the local and global minima using string method simulations, showing that the stabilities of the minima increase with increasing difference in the anisotropy of the ferromagnetic layers. We provide an analytical solution to the location of the minima in the energy landscape of coupled macrospins, which has good agreement with our micromagnetic results for a case involving ferromagnetic layers with the same thickness and anisotropy, no demagnetization field, and large exchange stiffness. These results are important to understand how best to employ noncollinear coupling in the next generation of thin film magnetic devices. 
\end{abstract}

\maketitle



\section{\label{sec:introduction}Introduction}

While magnetic computer memory dominated the nascent stages of fast-memory applications, it was quickly supplanted by solid-state, transistor-based, dynamic random-access memory (DRAM) and static random-access memory (SRAM)~\cite{chm_1970_2007}. Recently, however, there has been renewed academic and commercial interest in novel designs of magnetic RAM (MRAM), some of which promise to combine the density and low-cost of DRAM, the performance of static RAM (SRAM), and the non-volatility of hard disk drives~\cite{khvalkovskiy_basic_2013}. 

At the heart of contemporary MRAM designs is a trilayer configuration, which comprises two magnetic layers separated by a non-magnetic, insulating spacer. This is commonly referred to as a magnetic tunnel junction (MTJ). Typically, one layer has a \textit{maximized} uniaxial anisotropy (the hard layer), while the other has a \textit{reduced} but sufficiently large uniaxial anisotropy to maintain thermal stability (the soft layer)~\cite{mangin_current-induced_2006, weller_thermal_1999}. Writing operations are performed by reversing the relative direction of magnetization of these layers (either antiparallel or parallel) by switching the polarization of the softer layer, which is thus referred to as the ``free layer". The relative orientation of the layers' magnetization is measured through tunneling magnetoresistance. 

\begin{figure}[htb]
    \includegraphics[width=0.47\textwidth]{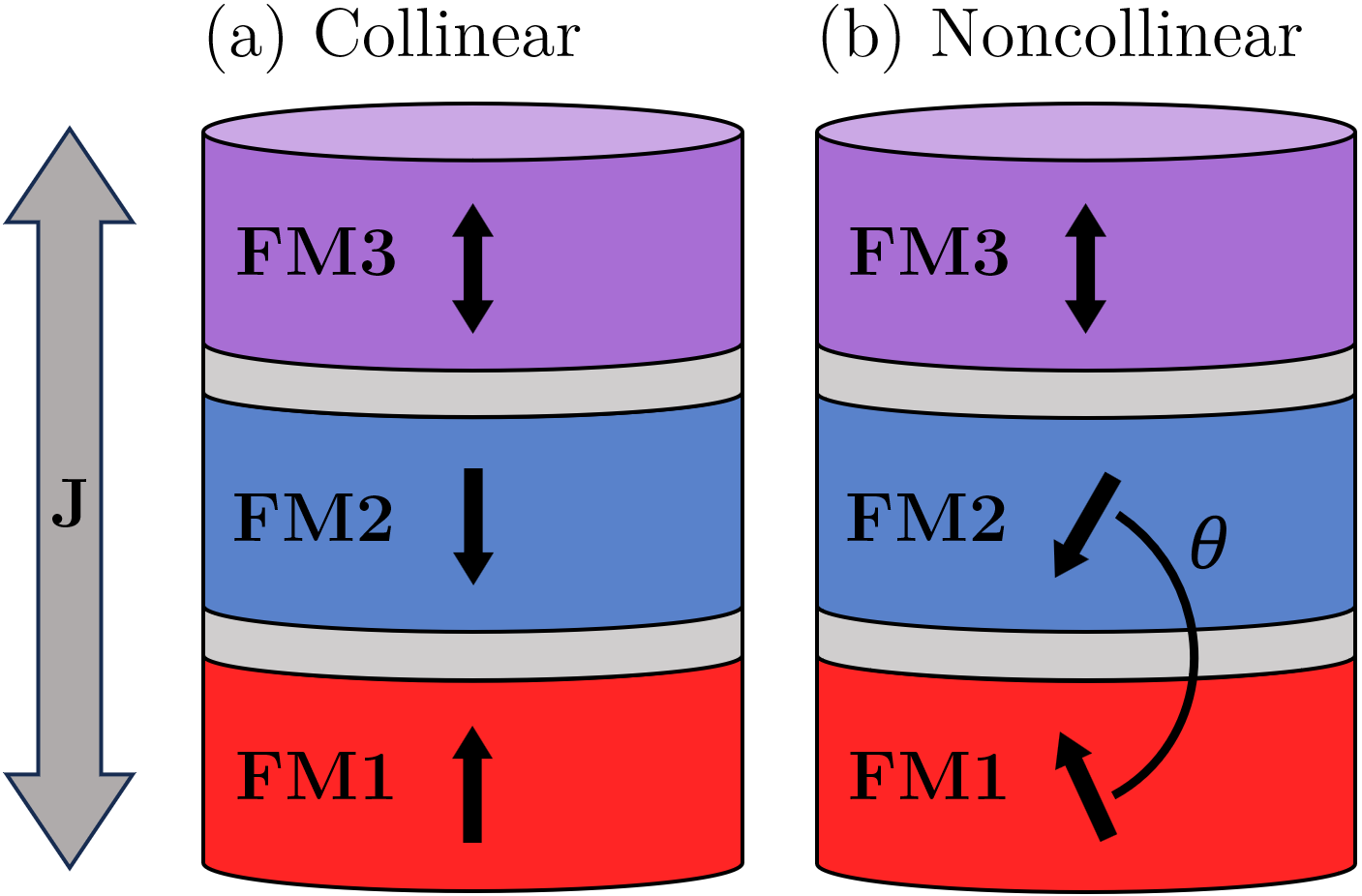}
    \caption{Schematic of an STT-MRAM stack, where a single bit of memory is stored in the relative orientation of the magnetic moment of the second and third ferromagnetic layers, FM2 and FM3. Writing is performed by reversing the direction of magnetization of FM3 by developing a current along either the positive or negative vertical axis. (a) A collinear STT-MRAM cell with $\theta = 180^\circ$ and (b) a noncollinear STT-MRAM cell, with $90^\circ < \theta < 180^\circ$.}
    \label{fig:STT-MRAMDiagramab}
\end{figure}

One such design of MRAM is especially exciting: spin-transfer torque MRAM (STT-MRAM), which allows the reading and writing operations of a single memory bit to be performed by the current supplied by a single driving MOSFET. During a write operation, a large current is developed across the perpendicularly magnetized magnetic layers, being spin-polarized by the hard layer. This current interacts with the magnetization of the free layer through spin-transfer torque (STT), wherein a spin-polarized current can induce magnetic reversal by exerting a torque on the local magnetic moment~\cite{slonczewski_current-driven_1996}. Thus, the direction of the applied current determines the resultant direction of magnetization in the free layer. In practice, STT-MRAM devices are often composed of three layers: the free layer (FM3), a reference layer (FM2), and a stabilizing layer (FM1), as shown in Fig.~\ref{fig:STT-MRAMDiagramab}(a). FM1 and FM2 are strongly antiferromagnetically exchange coupled, i.e. their moments are antiparallel. This coupling improves the stability of FM2 during writing processes and, ideally, eliminates unwanted stray fields on FM3~\cite{richter_how_2002, richter_simplified_2002}. In this work we will consider the design of FM1 and FM2 of a three-layer STT-MRAM structure, primarily with the assumption that FM3 is free and stores the magnetic bit. However, one can instead consider a design where both FM1 and FM2 comprise the free layer and store the magnetic information, while FM3 is the immobile hard layer. 

There is already widespread industry and academic interest in STT-MRAM~\cite{everspin_everspin_2008, chung_4gbit_2016, kultursay_evaluating_2013}, and while some modern designs are showing substantially reduced writing times~\cite{lee_fast_2020, safranski_reliable_2022}, switching currents and write-error-rates remain unacceptably large~\cite{kent_new_2015, worledge_write-error-rate_2023}. Reducing the switching current and increasing the writing speed while maintaining long-term thermal stability is of critical importance. It was demonstrated in \cite{sbiaa_magnetization_2013, matsumoto_spin-transfer-torque_2015} that the introduction of noncollinear coupling (an interlayer angle between $0^\circ$ and $180^\circ$) between the magnetic moments of FM1 and FM2 can substantially improve the performance of STT-MRAM devices in this regard by ensuring a non-zero (and controllable) spin-torque on FM3 at the beginning of switching processes, as shown in Fig.~\ref{fig:STT-MRAMDiagramab}(b). In devices without this modification, the only source of noncollinearity between FM2 and FM3 is the random off-axis thermal motion of the magnetic moments of the ferromagnetic layers. Instead, in this work we investigate the effect of biquadratic in addition to bilinear interlayer exchange coupling to create noncollinear magnetic alignment of two ferromagnetic layers. We primarily look from the perspective of application in STT-MRAM, but much of the results of this work can be applied to magnetic sensors, spin-torque oscillators, and other thin-film magnetic devices~\cite{nunn_control_2020}. 

In this article, we will first construct a macrospin model of the magnetic energy of two coupled ferromagnetic layers, FM1 and FM2 (Sec.~\ref{sec:theoryMacrospin}). This model will be used for analytical calculations of the magnetization state of this system. We will then present results from full micromagnetic simulations which illustrate the coupling angle between FM1 and FM2 as a function of the interlayer exchange coupling strengths and layer anisotropies of a nanopillar (Sec.~\ref{sec:couplingAngles}). Maintaining the same material parameters, we will then simulate the energy barriers between minima in the magnetic energy landscape of these layers (Sec.~\ref{sec:couplingBarriers}). In these simulations, we will consider the results only in terms of experimentally achievable material parameters. Finally, we will compare some of these simulated results to those from an analytic solution to the macrospin model, considering a simplified case for equal layer anisotropies and without the effects of the stray field (Sec.~\ref{sec:analyticSoln}). We will provide a discussion of our findings in the context of magnetic device design throughout. 


\section{\label{sec:theory}Theory}

\subsection{\label{sec:theoryCoupling}Interlayer exchange coupling}

The magnetization configuration of a multilayer is governed by multiple energy contributions. The areal interlayer exchange energy density~\cite{nunn_control_2020} represents the strength of coupling between layers of magnetic material separated by a non-magnetic spacer layer~\cite{stiles_interlayer_2005}. It can be expressed as
\begin{equation}\label{eq:Iexchange}
    E_\text{iex} = J_1\cos\left(\theta\right) + J_2\cos^2\left(\theta\right),
\end{equation}
where $J_1$ is the bilinear coupling constant, $J_2$ is the biquadratic coupling constant, and $\theta$ is the angle between the magnetic moments of the coupled layers. The bilinear term contributes energy minima at either $0^\circ$ (parallel) or $180^\circ$ (antiparallel) depending on the sign of $J_1$. In our convention, a positive $J_1$ favours antiferromagnetic coupling. Meanwhile, the biquadratic term is always positive and thus has identical minima at $90^\circ$ and $270^\circ$, which correspond to perpendicularly magnetized layers. When $J_2$ is zero or very small relative to $J_1$, the moments of the layers are collinear. Meanwhile, if $J_2 > J_1/2$ (for $J_1$ and $J_2 > 0$), and in the absence of other magnetic energy contributions, noncollinear configurations at the energy minimum will appear, as demonstrated by the first derivative of Eq.~(\ref{eq:Iexchange}).

It has been shown experimentally that the strength and sign of the bilinear term (determined by $J_1$) oscillates depending on the thickness and composition of the spacer~\cite{stiles_interlayer_2005}. The contribution of the biquadratic term ($J_2$) is shown to be induced by spatial variations in $J_1$~\cite{slonczewski_fluctuation_1991, nunn_non-collinear_2019}, the atomic surface roughness of the coupled layers~\cite{bland_ultrathin_2005}, pin-holes~\cite{bobo_pinholes_1999}, and loose spins~\cite{slonczewski_origin_1993}. Moreover, it was recently discovered that a new class of spacer layers containing a non-magnetic material (Ru or Ir) alloyed with a ferromagnetic material (Fe or Co) can be used to precisely control noncollinear coupling of magnetic moments, even those of smooth and uniform multilayers~\cite{nunn_control_2020, nunn_controlling_2023, besler_noncollinear_2023}. An atomistic model developed by Abert et al. in~\cite{abert_origin_2022} was shown to successfully reproduce the noncollinear coupling behaviour in these structures. These novel spacers allow the highly-scalable fabrication of noncollinearly coupled multilayers.

\subsection{\label{sec:theoryAnisotropy}Anisotropy energy}

The nanopillars simulated in this work consist of magnetic layers with uniform material parameters throughout the layer. Additionally, the ferromagnetic layers have uniaxial magnetocrystalline anisotropy perpendicular to the surface of the substrate. This anisotropy is larger than the shape anisotropy, forcing the magnetization of the films to orient perpendicular-to-the-plane. The volumetric uniaxial anisotropy energy density of a layer is given by
\begin{equation}\label{eq:EKuThin}
    \varepsilon_{K_u} = -K_u\cos^2(\phi),
\end{equation}
where $K_u$ is the uniaxial anisotropy constant and $\phi$ is the angle of the magnetic moment with respect to the anisotropy axis which, in this case, is perpendicular to the film plane.

In addition to crystallographic uniaxial anisotropy, the aspect ratio of a sample can impose a strongly preferred direction of magnetization, or shape anisotropy. The volumetric shape anisotropy energy density is defined as~\cite{arora_origin_2017}
\begin{equation}\label{eq:EKs}
    \varepsilon_{K_s} = N\frac{\mu_0}{2}M_s^2\cos^2\left(\phi\right),
\end{equation}
where $N$ is the demagnetization factor (a function of the sample aspect ratio), $\mu_0$ is the permeability of free space, and $\phi$ is, as defined in Eq.~(\ref{eq:EKuThin}), the angle between the magnetic moment and the sample surface normal.

The STT-MRAM nanopillars under study have an aspect ratio which confers a shape anisotropy causing magnetization in-plane. In this case, the combined effect of $E_{K_u}$ and $E_{K_s}$ determines the relaxed state of a sample---whichever is larger will dominate and determine whether a sample is magnetized in- or out-of-plane in its energy minimum. Thus, the effective volumetric anisotropy energy density is defined as
\begin{equation}\label{eq:EKEff}
    \varepsilon_{K_{\text{eff}}} = \varepsilon_{K_u} + \varepsilon_{K_s} = -K_{\text{eff}}\cos^2(\phi),
\end{equation}
with the effective anisotropy constant defined as
\begin{equation}\label{eq:EKEffsub}
    K_{\text{eff}} = K_u - N\frac{\mu_0}{2}M_s^2.
\end{equation}


\section{\label{sec:methods}Methods}

The simulations in this work are performed in \texttt{magnum.pi}~\cite{abert_magnumfe_2013}, a finite-element micromagnetic solver. In contrast to the analytical macrospin approach, introduced in Sec.~\ref{sec:theoryMacrospin}, the micromagnetic simulations in \texttt{magnum.pi} take into account the demagnetization field and allow for inhomogeneous magnetization configurations within the ferromagnetic layers~\cite{abert_micromagnetics_2019, suess_accurate_2023}. The demagnetization field includes the external interaction due to the stray fields generated by the ferromagnetic layers and the internal shape anisotropy. The boundary conditions used in the micromagnetic simulations are as defined in~\cite{abert_micromagnetics_2019, suess_accurate_2023}. \texttt{magnum.pi} requires a mesh of the simulation space: all numerically simulated structures in this work are a nanopillar with two ferromagnetic layers, FM1 and FM2. These have thicknesses of $d = \qty{3}{nm}$, while the nonmagnetic spacer has a thickness of $d = \qty{0.5}{nm}$. The entire stack is a circular cylinder with a radius of $\qty{15}{nm}$. This models the bottom two layers of Fig.~\ref{fig:STT-MRAMDiagramab}. Thus, the magnetic field generated by FM3 on FM1 and FM2 is not included in our calculations.

\subsection{\label{sec:methodsRelaxations}Micromagnetic relaxations}

To determine both local and global energy minima of the magnetization states of two interlayer exchange coupled ferromagnetic layers, each relaxation simulation is produced from one of two possible initial conditions: with the magnetic moments of FM1 and FM2 aligned collinearly parallel or collinearly antiparallel~\cite{lertzman-lepofsky_applications_2023}. In either case, their moments are initially oriented along the easy axes of the magnetization of FM1 and FM2, collinear with the z-axis. Herein, these initial collinear alignments will be referred to as ``PP" or ``AP", respectively. From these PP and AP states, the system is allowed to ``relax'' toward an energy minimum determined by the effective field contributions of the exchange effect, the anisotropy energy, demagnetization, and interlayer exchange coupling. This minimum determines the resultant interlayer angle. While the FM layers of the relaxed states may not always be collinear, the z-components of their magnetic moments will either point in the same or in opposite directions~\cite{lertzman-lepofsky_applications_2023}. Noncollinear states at an energy minimum will be referred to as \textit{noncollinear PP} (NCPP) and \textit{noncollinear AP} (NCAP) to reflect the alignment or antialignment of the z-components of the magnetization of FM1 and FM2. We note that PP initial conditions are most easily reproduced while fabricating film structures: the magnetization of a multilayer is biased by an external field along the easy axis of magnetization to align the magnetic moments of both FM layers, and then allowed to relax into an energy minimum. 

The anisotropy constant $K_u$ in each layer is chosen to take one of three values: $K_u = \qty{0.6}{\mega\joule\per\cubic\metre}$, $\qty{0.8}{\mega\joule\per\cubic\metre}$, and $\qty{1.0}{\mega\joule\per\cubic\metre}$. This ensures that $K_u > N\frac{\mu_0}{2}M^{2}{s}$ and that the thermal stability condition of $\frac{K_{1,2,\text{eff}} V}{K_B T} \geq 60$ is satisfied for the layer dimensions used in our calculations, corresponding to an operating temperature of $T = \qty{300}{\kelvin}$~\cite{mangin_current-induced_2006, weller_thermal_1999}. These anisotropy constants are representative of CoPt-based alloys and multilayers~\cite{shimatsu_high_2004, eyrich_effects_2014}. Both ferromagnetic layers have a saturation magnetization of $M_s = \qty{1}{\mega\ampere\per\meter}$ and an exchange stiffness of $A_{\text{ex}} = \qty{13}{\pico\joule\per\meter}$, which is likewise appropriate for CoPt films~\cite{eyrich_effects_2014}. These quantities define the exchange length, which determines the minimum size of the discretizations in the simulation mesh, which is thus chosen to be $\qty{3.0}{nm}$. We relax the system into a local minimum by integrating the Landau-Lifshitz-Gilbert (LLG) equation with a high Gilbert damping of $\alpha = 1.0$, which reduces computational run-times but does not impact the accuracy of the results.

In these simulations, we choose to vary $J_1$ over $\qtyrange{0}{4}{\milli\joule\per\square\metre}$ and $J_2$ over $\qtyrange{0}{3}{\milli\joule\per\square\metre}$ to reflect the range of values which are currently considered experimentally achievable across RuFe spacer layers~\cite{nunn_control_2020} while providing some additional context to act as a predictive tool for future work. 

\subsection{\label{sec:methodsString}String method}

\begin{figure}
    \includegraphics[width=86mm]{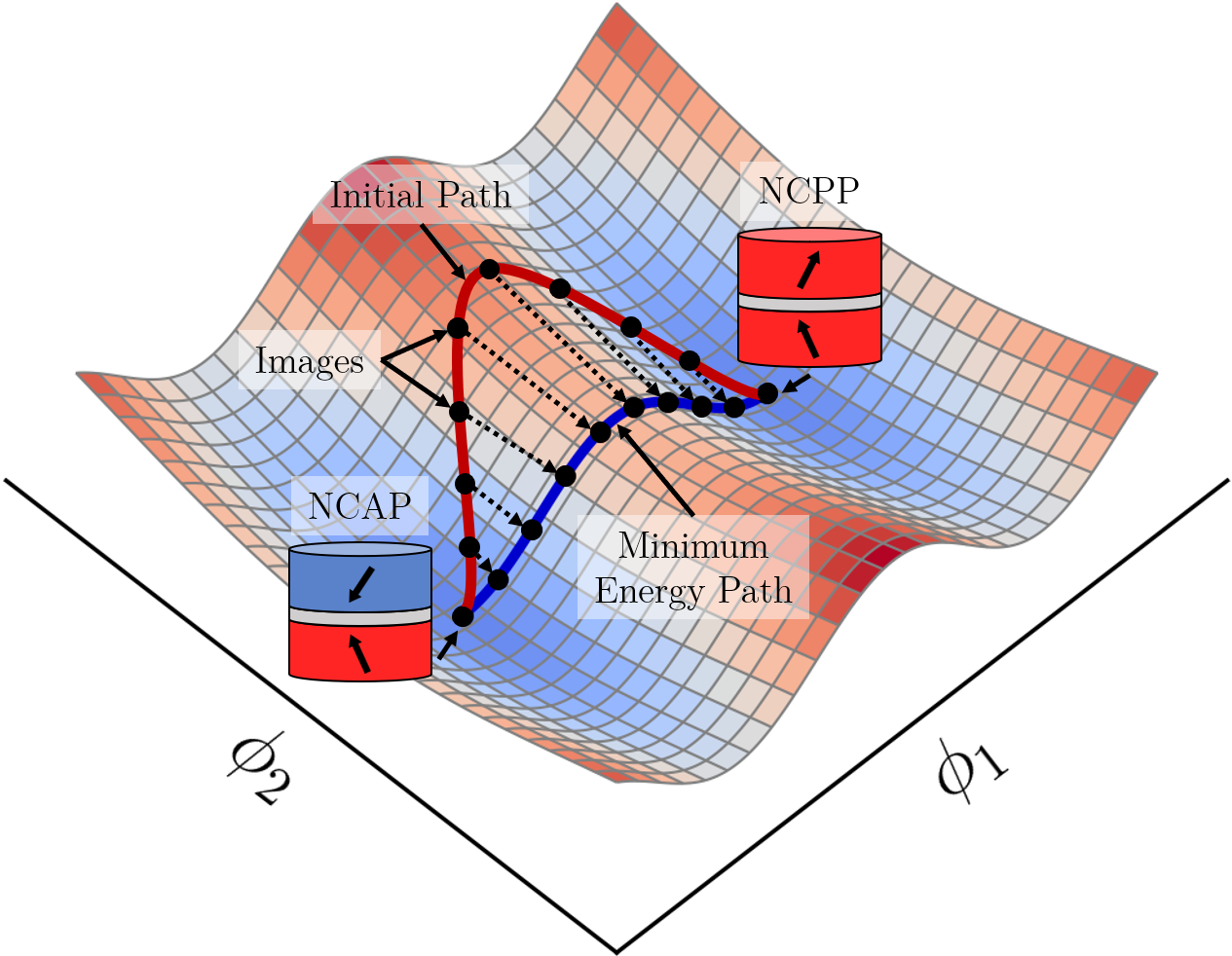}
    \caption{\label{fig:stringDiagramEdit}A visualization of the string method applied to the energy landscape defined by the energy dependence on the angles of the magnetic moments of FM1 and FM2 with respect to the z-axis, $\phi_1$ and $\phi_2$. Initially, the path between the two energy minima (NCPP and NCAP) of the relaxed FM layers is discretized along an arbitrary path into $n$ ``images" (initial path). Each of these states are iteratively relaxed using the steepest descent method and then rearranged equidistantly along an interpolated spline. After a sufficient number of iterations, this spline lies along the minimum energy path. 
    }
\end{figure}

Where the relaxation simulations described in Sec.~\ref{sec:methodsRelaxations} produce the same final state regardless of PP or AP initial condition, that state is known to be the global minimum for that pair of $J_1$, $J_2$ and $K_{u1}$, $K_{u2}$. Where these relaxations produce \textit{different} final alignments, this is indicative of a local minimum in the energy landscape. The minimum energy path between these minima can be computed using the string method, thereby identifying the local and global minimum and quantifying the height of the energy barrier between minima. String method simulations as implemented in \texttt{magnum.pi}~\cite{e_simplified_2007, koraltan_dependence_2020} are visualized in Fig.~\ref{fig:stringDiagramEdit}, wherein the path between each energy minimum is initially discretized into $n$ ``images" representing the transitional magnetization states between the NCPP and NCAP alignments. These images are iteratively relaxed toward the path with minimized barrier height using the steepest descent method, according to the direction of the damping term of the LLG equation, $\boldsymbol{m}\times(\boldsymbol{m}\times \boldsymbol{H}_{\text{eff}})$. The thermal stability of the minima can be determined from the difference in energy from each of the NCPP and NCAP states to the peak of the minimum energy path, which likewise allows us to identify the magnetic state of the global minimum. 
In this work, the energy path from an NCPP to an NCAP state is discretized using 20 images, which are evolved toward the minimum energy path over 100 iterations. 

\subsection{\label{sec:theoryMacrospin}Macrospin model}

\begin{figure}
    \includegraphics[width=86mm]{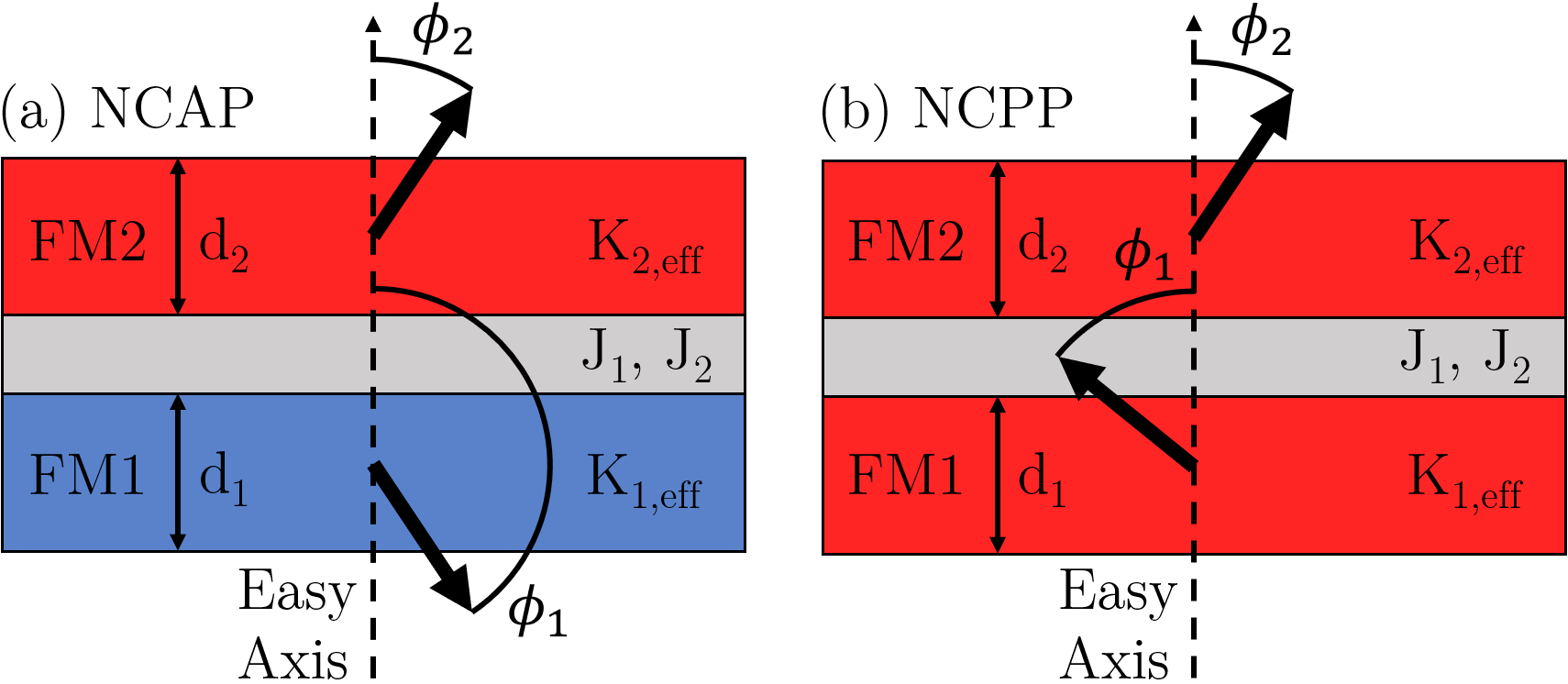}
    \caption[Labeled diagram for a macrospin model including uniaxial anisotropies.]{Schematic representation of a macrospin model of two ferromagnetic layers, FM1 and FM2, separated by a nonmagnetic spacer. $K_{1,\text{eff}}$ and $K_{2,\text{eff}}$ are effective anisotropies, $d_1$ and $d_2$ are layer thicknesses, and $J_1$ and $J_2$ are the interlayer exchange  coefficients for the coupling between FM1 and FM2. $\phi_1$ and $\phi_2$ are the angles formed by the macrospins with respect to the easy axis of magnetization. (a) A noncollinear antiparallel magnetization configuration (NCAP), and (b) a noncollinear parallel magnetization configuration (NCPP).}
    \label{fig:macrospinModel}
\end{figure}

A macrospin model provides a simplified yet effective framework for understanding the magnetostatic energy landscape in ferromagnetic layers. By assuming the exchange constant, $A_{\text{ex}}$, to be infinite, the model implies infinitely strong coupling between adjacent magnetic moments, ensuring that the magnetic moments within ferromagnetic layers are perfectly aligned. We likewise do not consider the effects of the stray fields produced by each ferromagnetic layer. This allows for the analysis to focus on the layer as a singular magnetic entity, without the complexity of inhomogeneous magnetic states. This assumption is justified for system sizes below the single-domain limit. We construct a model to represent the areal magnetic energy density of two ferromagnetic layers, separated by a nonmagnetic spacer. Fig.~\ref{fig:macrospinModel} illustrates the film structure considered in our macrospin model. The areal energy density for this configuration in the absence of an external magnetic field is given by
\begin{align}\label{eq:macro}
    E\left(\phi_1,\,\phi_2\right) &= -K_{1,\text{eff}}d_1\cos^2(\phi_1) - K_{2,\text{eff}}d_2\cos^2(\phi_2) \nonumber \\
      &\quad + J_1\cos(\phi_1 - \phi_2) + J_2\cos^2(\phi_1 - \phi_2),
\end{align}
where $K_{1,\text{eff}}$ and $K_{2,\text{eff}}$ are the effective anisotropy constants for ferromagnetic layers FM1 and FM2, respectively. The thicknesses of these layers are denoted by $d_1$ and $d_2$. The angles $\phi_1$ and $\phi_2$ are defined as the orientations of the magnetic moments in each layer relative to the easy axis of magnetization, as shown in Fig.~\ref{fig:macrospinModel}. $J_1$ and $J_2$ are the bilinear and biquadratic interlayer coupling strengths, respectively. 

\section{\label{sec:results}Results}

\subsection{\label{sec:couplingAngles}Micromagnetic interlayer coupling angles}

\begin{figure*}[ht]
    \includegraphics[width=179mm]{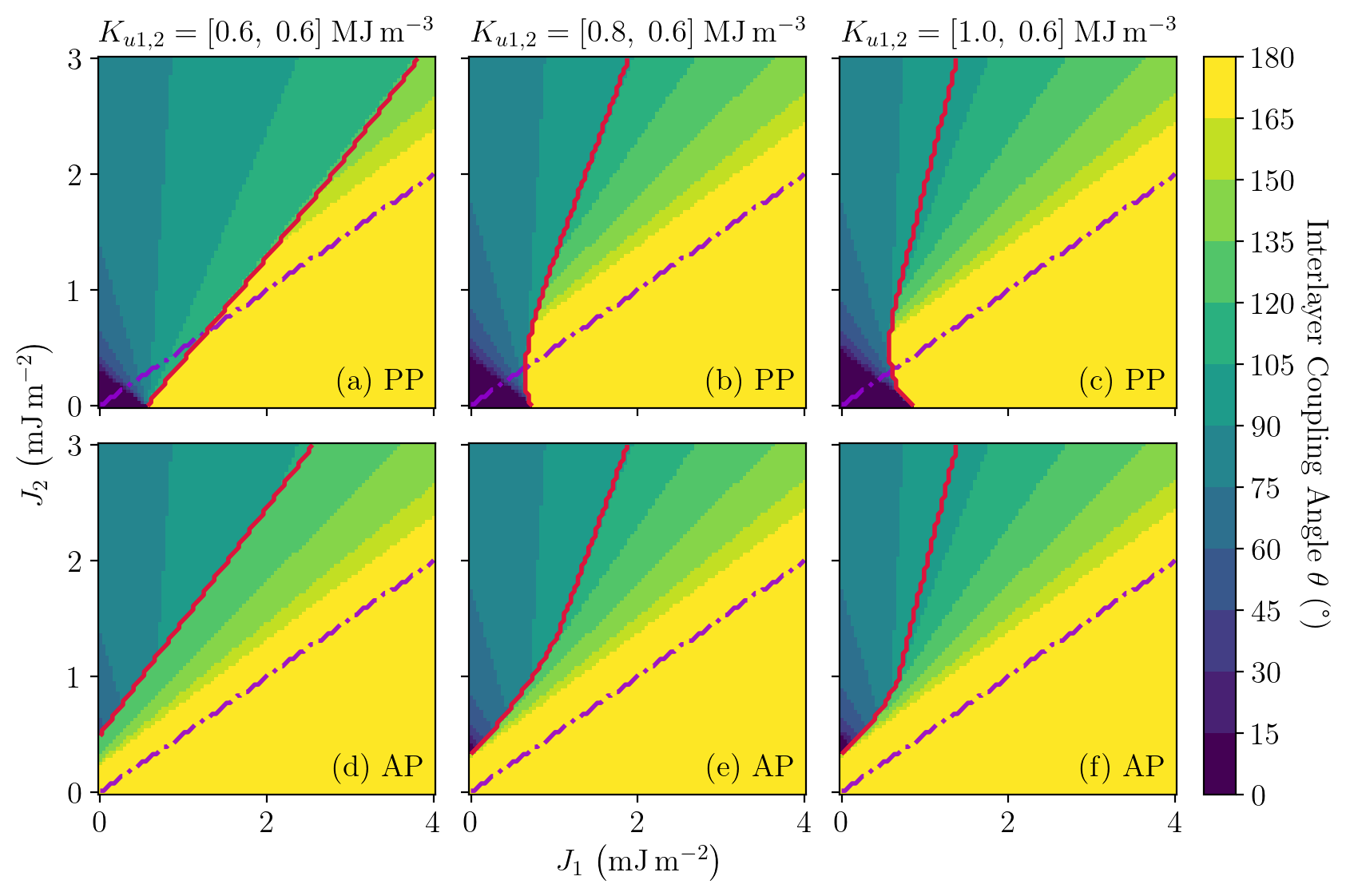}
    \centering
    \caption{\label{fig:interlayerCouplingAngles}Simulated interlayer coupling angle for relaxations starting from PP, (a) to (c), and AP initial conditions, (d) to (f), for varying interlayer coupling constants, $J_1$ and $J_2$, and uniaxial anisotropies in FM1, $K_{u1} = \qty{0.6}{\mega\joule\per\cubic\meter}$, $\qty{0.8}{\mega\joule\per\cubic\meter}$, and $\qty{1.0}{\mega\joule\per\cubic\meter}$. FM2 is the same in all plots, $K_{u2} = \qty{0.6}{\mega\joule\per\cubic\meter}$. The red line demarcates the final state: left of the line for NCPP and PP, right for NCAP and AP. The dashed purple line inscribes the $J_2 > J_1/2$ condition for noncollinearity taking into account only the interlayer exchange coupling energy.}
\end{figure*}
The results of micromagnetic simulations of the angle between the magnetic moments of FM1 and FM2, the interlayer coupling angle, are shown in Fig.~\ref{fig:interlayerCouplingAngles}, for varying coupling, anisotropies, and either parallel (PP) or antiparallel (AP) initial conditions. The interlayer coupling angle at the energy minimum is given according to the colourbar. Multilayers that relax to an NCPP state have values of $J_1$ and $J_2$ left of the red line. Conversely, multilayers that relax to an NCAP state have values of $J_1$ and $J_2$ right of the red line. The dashed purple line shows the $J_2 > J_1/2$ condition for noncollinearity. Taking into account only interlayer exchange coupling, multilayers with $J_1$ and $J_2$ above and to the left of this line will favour a noncollinear alignment of magnetic moments of FM1 and FM2. The addition of anisotropy to the model shifts the required value of $J_2$ to produce noncollinearity larger than $J_1/2$. For use as the pinning and reference layers of a noncollinearly coupled STT-MRAM device, one desires an NCAP multilayer, which minimizes the stray field on FM3. This configuration must exist for values of $J_1$ and $J_2$ which can be experimentally obtained from the coupling of FM1 and FM2 across RuFe and IrFe spacer layers~\cite{nunn_control_2020, besler_noncollinear_2023}.

For multilayers with the same uniaxial anisotropies, as in Figs.~\ref{fig:interlayerCouplingAngles}(a) and (d), there is a marked difference in the size and shape of the noncollinear regions produced by AP and PP initial conditions. This is indicative of the existence of two energy minima in the region where Fig.~\ref{fig:interlayerCouplingAngles}(a) and (d) differ. The energy states in these minima, either NCPP or NCAP, resemble the macrospin configurations in Fig.~\ref{fig:macrospinModel}. As well, there does not exist a large region of NCAP states with values of $J_1$ and $J_2$ which are produced by current spacer layers and is invariant through changes in the initial condition: only the area between the red line and the yellow region in Fig.~\ref{fig:interlayerCouplingAngles}(a) satisfies this condition. Therefore, multilayers with equal anisotropies may not be appropriate for use as a reference layer in STT-MRAM. The regions where the relaxed state depends on the initial conditions (where two minima exist) are shown in Fig.~\ref{fig:ppapCompare}. Meanwhile, Figs.~\ref{fig:interlayerCouplingAngles}(b, c, e, f) show interlayer angles for multilayers with different anisotropies. These show a clear trend in comparison to Figs.~\ref{fig:interlayerCouplingAngles}(a) and (d): the region where the noncollinear states are independent of the initial conditions is much larger and extends to much lower values of $J_1$ and $J_2$. This region is shown to grow with an increasing difference between the anisotropies of FM1 and FM2. Indeed, for $K_{u1,2} = [1.0,\; 0.6]\; \unit{\mega\joule\per\cubic\meter}$, the desired NCAP states are completely invariant to initial conditions for coupling values as low as $J_1 \approx \qty{0.59}{\milli\joule\per\square\meter}$ and $J_2 \approx \qty{0.63}{\milli\joule\per\square\meter}$, which demonstrates that samples with these material parameters should be reproducible in experiment with currently available spacer layers. 

We are confident of the presence of at most two minima in this energy landscape. Preliminary simulations were performed with a randomized initial orientation of each magnetic spin. The relaxed states produced by these simulations were entirely consistent with the ordered PP and AP initial conditions of the presented results. 

\subsection{\label{sec:couplingBarriers}Micromagnetic energy barriers}

\begin{figure}[ht]
    \includegraphics[width=84mm]{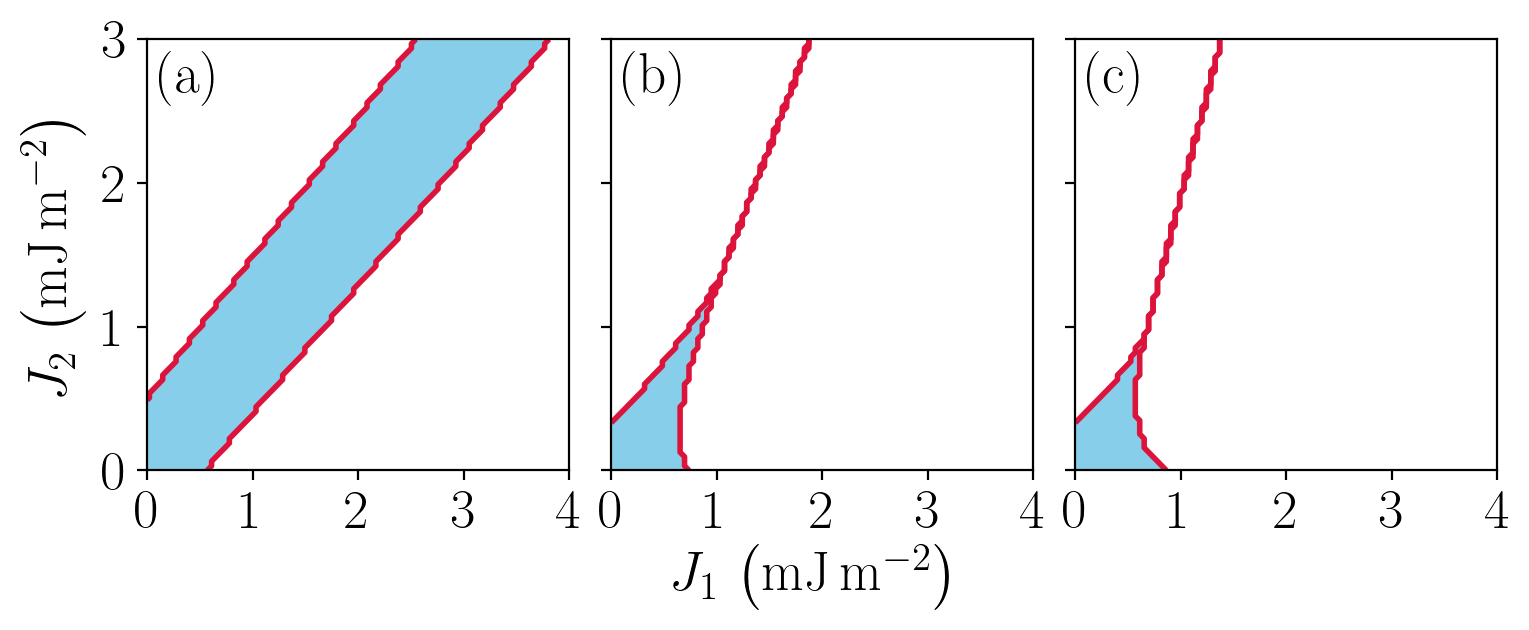}
    \caption{\label{fig:ppapCompare}
    The regions of the relaxation simulations for which two energy minima exist are enclosed by the red lines and coloured light blue. The boundaries are calculated in Fig.~\ref{fig:interlayerCouplingAngles}, and they are defined by the transition from NCPP to NCAP for a given initial condition and pair of anisotropies. The layers have anisotropies (a) $K_{u1,2} = [0.6,\; 0.6]\; \unit{\mega\joule\per\cubic\metre}$, (b) $K_{u1,2} = [0.8,\; 0.6]\; \unit{\mega\joule\per\cubic\metre}$, and (c) $K_{u1,2} = [1.0,\; 0.6]\; \unit{\mega\joule\per\cubic\metre}$.}
\end{figure}
As shown in Sec.~\ref{sec:couplingAngles}, there exist two energy minima (NCPP or NCAP) for certain values of $J_1$ and $J_2$ and $K_{u1,2}$. It has likewise been shown that the range of $J_1$ and $J_2$ that produce both minima is larger when the multilayered structures have layers with equal anisotropies. For given anisotropies of FM1 and FM2, this region is defined by the area between the red lines from each initial condition in Fig.~\ref{fig:interlayerCouplingAngles}. For example, the blue region of Fig.~\ref{fig:ppapCompare}(a) is obtained from the area between the red lines calculated from PP and AP initial conditions in Fig.~\ref{fig:interlayerCouplingAngles}(a) and (d), respectively. These regions define the pairs of $J_1$ and $J_2$ for which we perform string method simulations. This is by necessity, since the string method requires at least one stable state for each end of the path. 

\begin{figure}[ht]
    \includegraphics[width=84mm]{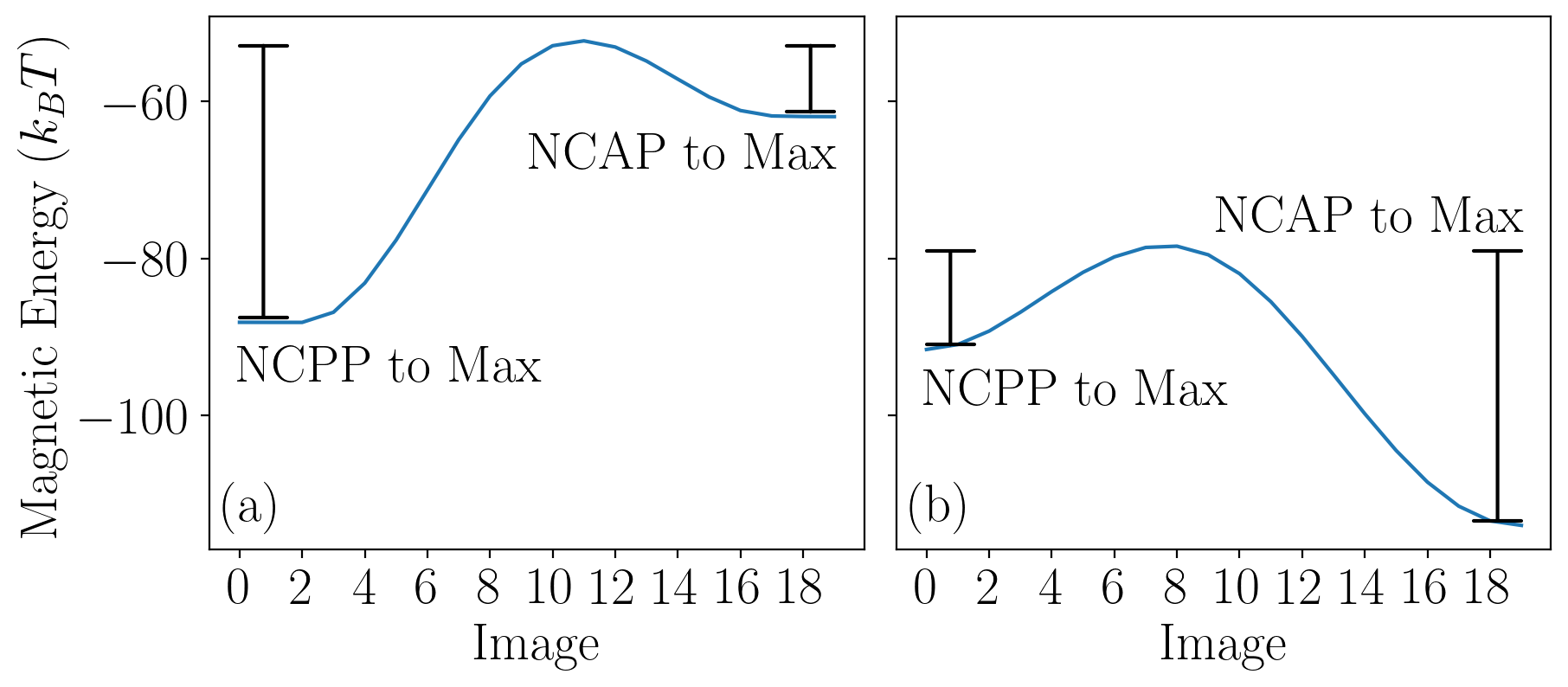}
    \caption{\label{fig:labeledBarriers}Minimum energy path and energy barriers calculated using the string method for (a) $J_1 = \qty{0.80}{\milli\joule\per\square\meter}$, $J_2 = \qty{0.57}{\milli\joule\per\square\meter}$ and (b)
    $J_1 = \qty{0.55}{\milli\joule\per\square\meter}$, $J_2 = \qty{0.66}{\milli\joule\per\square\meter}$, with $K_{u1,2} = [0.6,\; 0.6]\; \unit{\mega\joule\per\cubic\metre}$. The NCAP state is the global minimum in (a), while the NCPP state is the global minimum in (b). The barrier height is measured with respect to each minimum. Each ``image'' corresponds to a black dot along the minimum energy path in the string method diagram, Fig.~\ref{fig:stringDiagramEdit}.}
\end{figure}

\begin{figure*}[ht]
    \includegraphics[width=179mm]{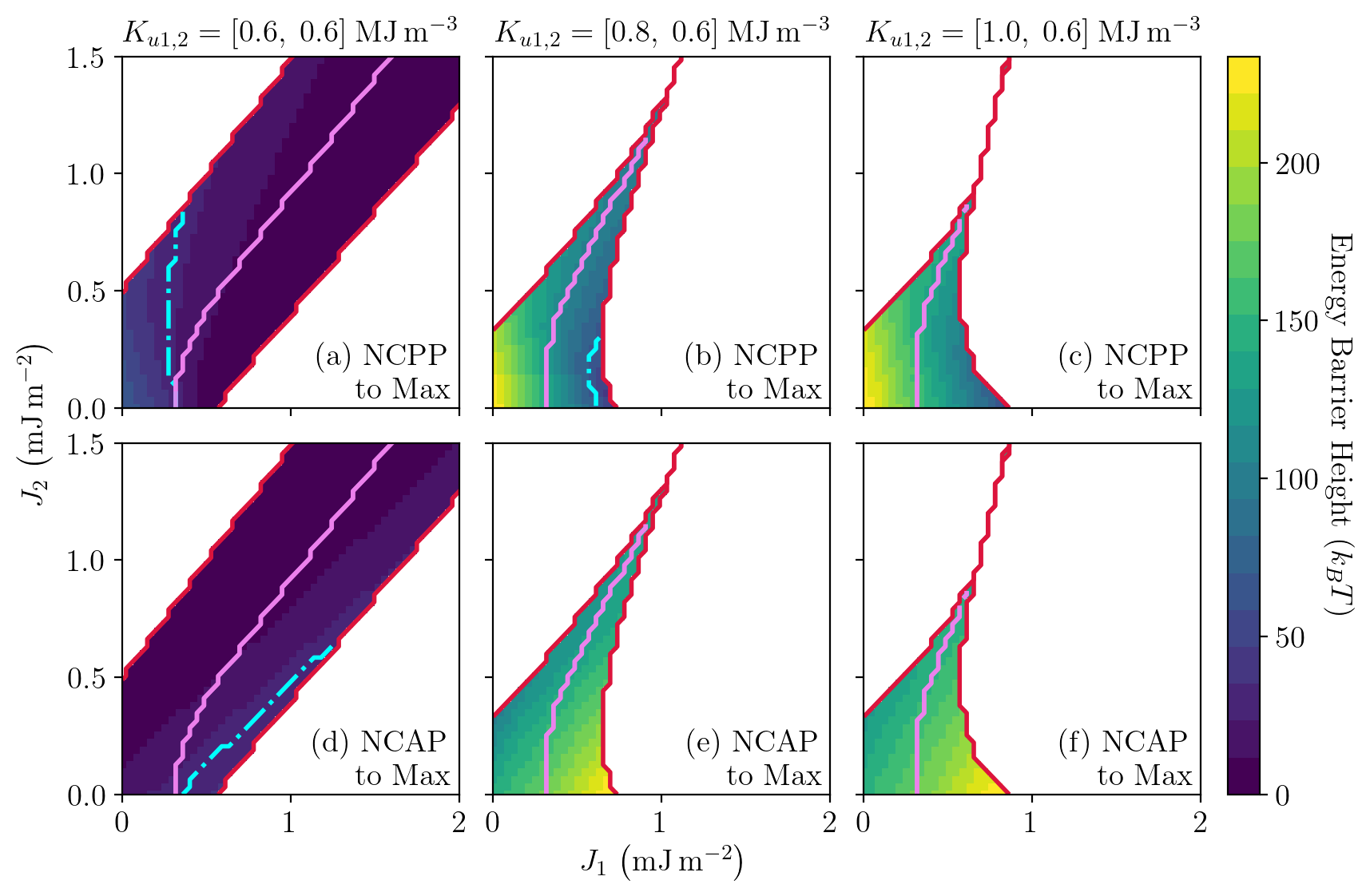}
    \caption{\label{fig:interlayerCouplingBarriers}Energy barrier heights between relaxed NCPP and NCAP states with varying uniaxial anisotropies, simulated for values of $J_1$ and $J_2$ for which there are two minima (the blue regions of Fig.~\ref{fig:ppapCompare}, zoomed to highlight a smaller range of values of $J_1$ and $J_2$). Plots (a) to (c) show the height of the barrier from the NCPP state to the energy maximum, while (d) to (f) show the height of the barrier from NCAP to the maximum. The violet line divides the two regions with differing magnetic alignments at the global energy minimum: above and to the left for NCPP, below and to the right for NCAP. The dashed cyan line divides regions with barrier heights above and below $\qty{60}{k_BT}$.}
\end{figure*}

Fig.~\ref{fig:labeledBarriers} shows two evolutions from an NCPP to an NCAP state: (a) where NCPP is the global minimum and (b) where NCAP is the global minimum, obtained from string method simulations. These correspond  to multilayers with $J_1$ and $J_2$ above and below the violet lines of Fig.~\ref{fig:interlayerCouplingBarriers}(a) and (d), respectively. As in the figure, one can define \textit{two} barrier heights for a given pair of $J_1$ and $J_2$, one from NCPP to the maximum barrier energy and another from NCAP to the maximum barrier energy. Each describes the stability of its respective minimum under thermal fluctuations or other sources of energy. 

The results of string method simulations for the values of $J_1$ and $J_2$ that produce local minima, one for each of the plots in Fig.~\ref{fig:ppapCompare}, are shown in Fig.~\ref{fig:interlayerCouplingBarriers}. This figure is zoomed-in with respect to Fig.~\ref{fig:interlayerCouplingAngles} and Fig.~\ref{fig:ppapCompare} to highlight lower values of $J_1$ and $J_2$. As in Fig.~\ref{fig:interlayerCouplingAngles}, the results in Fig.~\ref{fig:interlayerCouplingBarriers} are given in two plots for each pair of anisotropies, $K_{u1,2} = [0.6,\; 0.6]\; \unit{\mega\joule\per\cubic\metre}$, $K_{u1,2} = [0.8,\; 0.6]\; \unit{\mega\joule\per\cubic\metre}$, and $K_{u1,2} = [1.0,\; 0.6]\; \unit{\mega\joule\per\cubic\metre}$. The figure shows the difference in energy from each NCPP state to the maximum barrier energy (Fig.~\ref{fig:interlayerCouplingBarriers}(a)-(c)), and the difference in energy from each NCAP state to the maximum barrier energy (Fig.~\ref{fig:interlayerCouplingBarriers}(d)-(f)). Thus, the plot is coloured according to the magnitude of the energy barriers defined in Fig.~\ref{fig:labeledBarriers}. These results show that there exist critical values of $J_1$ and $J_2$ that separate the region where NCPP is the global energy minimum and where NCAP is the global energy minimum, which is demarcated by violet lines in the figures. These lines bend down toward the $J_1$ axis for small $J_2$ due to the increasing relative effect of the demagnetization field for low coupling energy density. 

For the studied anisotropies, barrier heights are larger for multilayers with unequal anisotropies as compared with those with equal anisotropies for the same values of $J_1$ and $J_2$. This indicates that both the NCAP and NCPP states are more stable when anisotropies are unequal, as demonstrated by the trend over Fig.~\ref{fig:interlayerCouplingBarriers}(a-c, d-f). Fig.~\ref{fig:interlayerCouplingBarriers}(c) shows that the energy barrier for every state is above $\qty{60}{k_BT}$ for $K_{u1,2} = [1.0,\; 0.6]\; \unit{\mega\joule\per\cubic\meter}$. Similarly, Fig.~\ref{fig:interlayerCouplingBarriers}(b) shows that barriers heights are below $\qty{60}{k_BT}$ only in a very small region for $K_{u1,2} = [0.8,\; 0.6]\; \unit{\mega\joule\per\cubic\meter}$. In contrast, for $K_{u1,2} = [0.6,\; 0.6]\; \unit{\mega\joule\per\cubic\meter}$  (Fig.~\ref{fig:interlayerCouplingBarriers}(a)), the majority of the barriers are below $\qty{60}{k_BT}$. According to~\cite{weller_thermal_1999}, a barrier height of at least $\qty{60}{k_BT}$ is required to guarantee stability of a magnetic state to thermal fluctuations for at least 10 years. This indicates that there is a high probability that layers with $K_{u1,2} = [0.6,\; 0.6]\; \unit{\mega\joule\per\cubic\meter}$ will repeatedly transition between two minima for most values of $J_1$ and $J_2$ in the long term. However, these barrier heights could potentially be tuned through adjustments to layer dimensions, allowing the structures in the two-minima band of Fig.~\ref{fig:interlayerCouplingBarriers}(a) to be used in a probabilistic computing device~\cite{grollier_neuromorphic_2020}. 

We have focused thus far on applications in a three-layer STT-MRAM device, where FM1 and FM2 comprise a stable reference magnetization for a free FM3. Alternatively, FM1 and FM2 can be designed to comprise the memory storage element themselves, with FM3 used as a stable reference. In this case, it is \textit{required} that there exist two stable magnetization states in the energy landscape, both NCPP and NCPP, to store magnetic information. That is, these devices can only exist within the blue regions of Fig.~\ref{fig:ppapCompare}. However, such a device will, in general, have an unequal energy barrier from each minimum to the peak of the barrier, as represented by the states in Fig.~\ref{fig:labeledBarriers}. This results in an anisotropy in the direction of the writing current, which may be desirable in some applications. In contrast, in order for a noncollinear FM1 and FM2 to have symmetric energy barriers, $J_1$ and $J_2$ must lie precisely along the violet line of Fig.~\ref{fig:interlayerCouplingBarriers}. This would pose significant fabrication challenges owing to the degree of specificity of the values of $J_1$ and $J_2$. 

\subsection{\label{sec:analyticSoln}Analytical solution to the macrospin model}

The macrospin model, as outlined in Eq.~(\ref{eq:macro}), can be  analytically solved for the case of ferromagnetic layers having the same anisotropy ($K_{1,\text{eff}}=K_{2,\text{eff}}=K_u$), and thicknesses ($d_1 = d_2 = d$). Under these conditions, the areal energy density can be expressed as:
\begin{align}\label{eq:macro2}
    E(\phi_1, \phi_2) &= -K_u d\cos^2(\phi_1) - K_ud\cos^2(\phi_2) \nonumber \\
    &\phantom{=} + J_1\cos(\phi_1 - \phi_2) + J_2\cos^2(\phi_1 - \phi_2).
\end{align}
To identify the relative minima within this energy landscape, we employ the second partial derivative test. First, we find the critical points of $E(\phi_1, \phi_2)$ by setting its first partial derivatives with respect to $\phi_1$ and $\phi_2$ to zero. This yields a system of two equations, and their summation leads to a condition,
\begin{equation}
    \sin\left(2\phi_1\right) + \sin\left(2\phi_2\right) = 0.
\end{equation}
We restrict the analysis to $0 \leq \phi_{1,2} \leq 2\pi$. The unique solutions identified are $\phi_2 = \pi - \phi_1$ and $\phi_2 = - \phi_1$.

\begin{figure}[ht]
    \includegraphics[width=84mm]{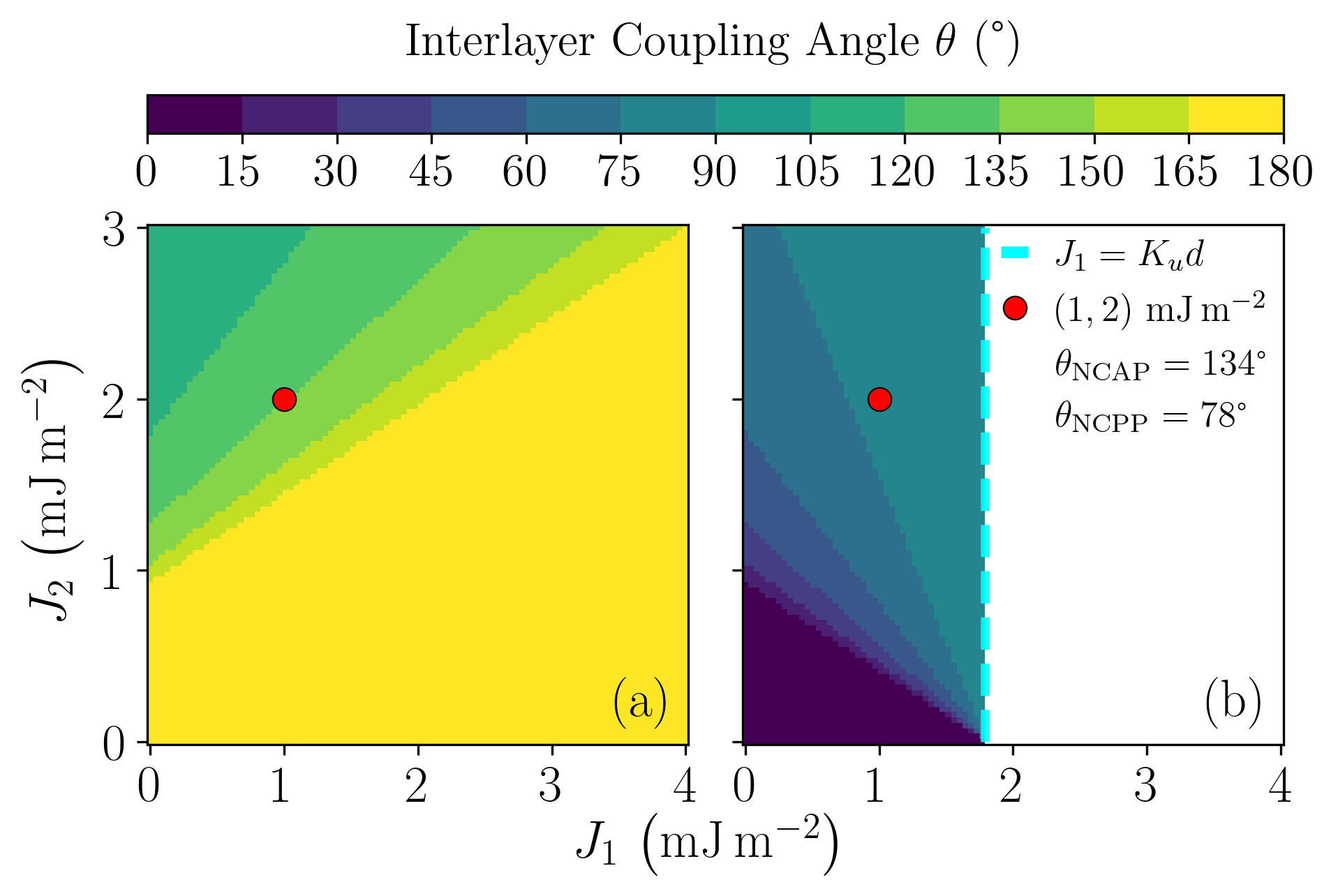} 
    \caption{Variation in the interlayer coupling angle between two ferromagnetic layers as predicted by the macrospin model (Eq.~(\ref{eq:macro2})). (a) The first solution, as described by Eq.~(\ref{eq:firstSol}) when $J_1 + K_u d < 2J_2$, and $\theta=180^{\circ}$ when $J_1 + K_u d > 2J_2$. (b) The second solution, in accordance with Eq.~(\ref{eq:secondSol}) when $K_u d < J_1 + 2J_2$ and $ J_1 < K_u d$, and $\theta=0^{\circ}$ when $K_u d > J_1 + 2J_2$. The region where both solutions exist is demarcated by a dashed violet line, indicating the condition $J_1<K_u d$. The solid red circle on each plot highlights a specific pair of $J_1$ and $J_2$ values, $J_1=\qty{1.0}{\milli\joule\per\square\metre}$ and $J_2=\qty{2.0}{\milli\joule\per\square\metre}$, for which there exist two minima at $\theta_{\text{NCAP}}=134^{\circ}$ and $\theta_{\text{NCPP}}=78^{\circ}$. Both layers have identical magnetic anisotropies and thicknesses, with $K_{u1,2} = \qty{0.6}{\mega\joule\per\cubic\metre}$ and $d_{1,2} = \qty{3.0}{nm}$, respectively.}
    \label{fig:macrospinCompare}
\end{figure}

Next, we compute the second partial derivatives with respect to $\phi_1$ and $\phi_2$, as well as the mixed partial derivative. We can then evaluate the Hessian matrix at the two determined critical points. A critical point is identified as a minimum if the determinant of the Hessian matrix is positive and the second derivative with respect to either $\phi_1$ or $\phi_2$ is also positive. These solutions represent the macrospin orientations wherein the energy function $E(\phi_1, \phi_2)$ attains a relative minimum. Furthermore, we restrict $J_{1,2} > 0$, $K_u > 0$, and $d > 0$. Fig.~\ref{fig:macrospinCompare} shows the interlayer angle produced from two distinct solutions to Eq.~(\ref{eq:macro2}) for $K_{u1,2} = \qty{0.6}{\mega\joule\per\cubic\metre}$ and $d_{1,2} = \qty{3.0}{nm}$. 

The first solution is given by the following equations:
\begin{align}
    \begin{split}\label{eq:macrospinSolution1}
        &\phi_1 = \pi - \arctan\left(\sqrt{-\frac{J_1 - 2 J_2 + K_u d}{J_1 + 2 J_2 + K_u d}}\right), \\ 
        &\phi_2 = \pi - \phi_1,
    \end{split}
\intertext{where \( J_1 + K_u d < 2J_2 \), which gives interlayer angle $\theta_{\text{NCAP}} = \phi_1 - \phi_2$:}
    \begin{split}
    &\theta_{\text{NCAP}} = \pi - 2\arctan\left(\sqrt{-\frac{J_1 - 2 J_2 + K_u d}{J_1 + 2 J_2 + K_u d}}\right).
    \label{eq:firstSol}
    \end{split}
\intertext{This angle is plotted as a function of $J_1$ and $J_2$ in Fig.~\ref{fig:macrospinCompare}(a). When \( J_1 + K_ud > 2J_2\), the interlayer angle is $\pi$, which is represented by the yellow area in the figure. The second solution is given by:}
    \begin{split}\label{eq:macrospinSolution2}
    &\phi_1 = \arctan\left(\sqrt{\frac{J_1 + 2J_2 - K_u d}{-J_1 + 2J_2 + K_u d}}\right), \\
    &\phi_2 = - \phi_1,
    \end{split}
\intertext{where \(K_u d < J_1 + 2J_2 \) and \( J_1 < K_u d \), which gives interlayer angle $\theta_{\text{NCPP}} = \phi_1 - \phi_2$:} 
    \begin{split}
     &\theta_{\text{NCPP}} = 2\arctan\left(\sqrt{\frac{J_1 + 2J_2 - K_u d}{-J_1 + 2J_2 + K_u d}}\right).
    \label{eq:secondSol}
    \end{split}
\end{align}
This angle is plotted as a function of $J_1$ and $J_2$ in Fig.~\ref{fig:macrospinCompare}(b). When \(K_u d > J_1 + 2J_2 \), the interlayer angle is $0$, which is represented by the purple area in the figure. Thus, the presence of a second minimum is contingent upon satisfying the condition $J_1 < K_u d$. This finding highlights the essential influence of the ratio between $J_1$ and $K_u d$ on the system's energy landscape. For an arbitrary point ($J_1=\qty{1.0}{\milli\joule\per\square\metre}$ and $J_2=\qty{2.0}{\milli\joule\per\square\metre}$), that satisfies this condition ($J_1 < K_u d$), there are two distinct minima: $\theta_{\text{NCAP}}=134^{\circ}$ and $\theta_{\text{NCPP}}=78^{\circ}$, as illustrated in Fig.~\ref{fig:macrospinCompare}.
\begin{figure}[ht]
    \includegraphics[width=84mm]{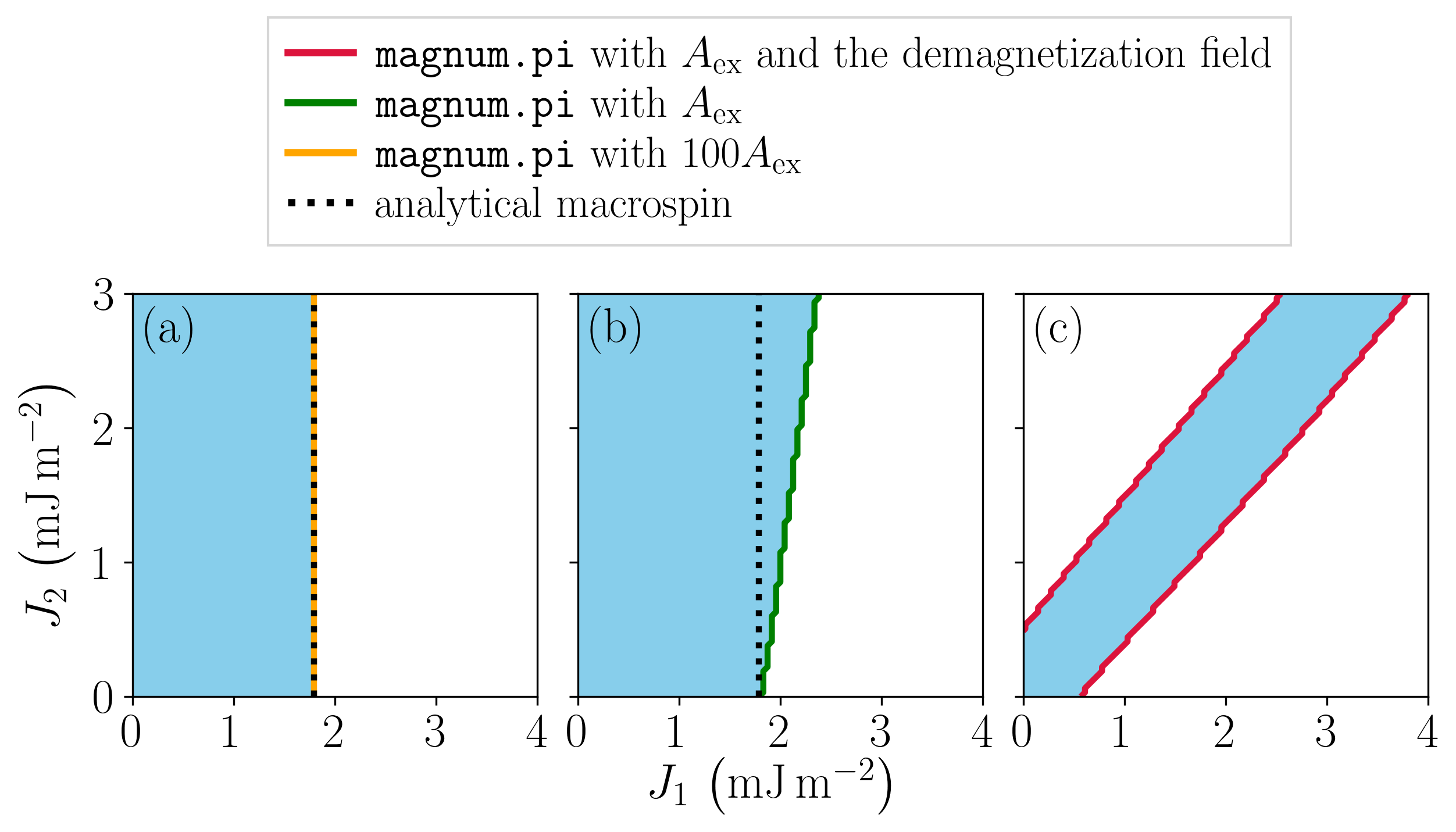} 
    \caption{Visualization of regions exhibiting dual energy minima, highlighted in light blue. (a) The macrospin calculations and simulations agree well when the exchange stiffness in the simulation is set to $100A_{\text{ex}}$ ($A_{\text{ex}} = \qty{13}{\pico\joule\per\m}$) and the demagnetization field is not considered. (b) The expansion of the two minima region obtained from the simulation using exchange stiffness $A_{\text{ex}}$ and neglecting the demagnetization field. (c) With the introduction of the demagnetization field in the simulation, the region tilts to the right, leaving an area with one minimum in the top left. Both layers have identical magnetic anisotropies and thicknesses, with $K_{u1,2} = \qty{0.6}{\mega\joule\per\cubic\metre}$ and $d_{1,2} = \qty{3.0}{nm}$, respectively.}     
    \label{fig:macrospinCompare2}
\end{figure}

In Fig.~\ref{fig:macrospinCompare2}(a), we show that the region exhibiting dual minima in the analytical macrospin solution is consistent with the results obtained from micromagnetic simulations that neglect the demagnetization field and use an exchange stiffness that is 100 times larger than in the earlier simulations ($A_{\text{ex}} = \qty{13}{\pico\joule\per\meter}$). This ensures a coherent rotation of magnetization in the simulation, mirroring the behavior presupposed in the macrospin model. Fig.~\ref{fig:macrospinCompare2}(b) shows that the range of $J_1$ and $J_2$ values for which two minima exist expands when $A_{\text{ex}} = \qty{13}{\pico\joule\per\meter}$ is used. In a simulation with this exchange stiffness that includes the demagnetization field, this region tilts to the right, leaving an area in the top left with only one minimum, as shown in Fig.~\ref{fig:macrospinCompare2}(c). This shift is due to the effects of the stray field, which promotes the parallel alignment of the magnetic moments of the ferromagnetic layers. 


\section{\label{sec:discussion}Discussion}

This paper presents a study of the effects of interlayer exchange coupling and uniaxial anisotropy strengths on the magnetic energy landscape of two coupled ferromagnetic layers. We showed that, when relaxing from a collinear parallel or antiparallel state, the moments of the ferromagnetic layers of a nanopillar fall into one of two energy minima: noncollinear parallel (NCPP) or noncollinear antiparallel (NCAP), depending on coupling strength and the magnitude of the difference in the layer uniaxial anisotropies. For certain material parameters, the relaxed state is entirely determined by initial conditions, due to the presence of a local minimum in the energy landscape. We showed that the range of values of experimentally reproducible $J_1$ and $J_2$~\cite{mckinnon_thermally_2022, nunn_control_2020}, and for which the relaxed state is NCAP, is \textit{maximized} when the difference in anisotropies is large. Additionally, we demonstrated that the range of $J_1$ and $J_2$ which produce two minima is minimized in these structures. A large parameter space with NCAP states is critical for use in STT-MRAM with a noncollinear reference layer. Furthermore, string method simulations showed that the energy barriers between NCPP and NCAP states (when both are present) are mostly smaller than the $\qty{60}{k_BT}$ requirement for stability over 10 years when $K_{u1,2} = [0.6,\; 0.6]\; \unit{\mega\joule\per\cubic\meter}$. Conversely, the vast majority of the energy barriers are larger than this threshold when layer anisotropies are unequal. These string method simulations also showed the regions for which NCAP or NCPP is the global energy minimum state. Furthermore, an analytic solution to a macrospin model of interlayer exchange coupling for equal uniaxial anisotropies and ferromagnetic layer thicknesses showed very good agreement with the micromagnetic simulations, without considering the effects of the demagnetization field and assuming large exchange stiffness. We identified an analytic expression for the condition which produces two minima in this case: $J_1 < K_u d$. These results provide targets and constraints on the interlayer exchange coupling strengths and layer anisotropies required by noncollinear STT-MRAM designs and other thin film magnetic devices.


\begin{acknowledgments}
The simulations presented in this work were enabled in part by computational resources managed and supported by the SFU Research Computing Group (\url{rcg.sfu.ca}) and the Digital Research Alliance of Canada (\url{alliancecan.ca}). This research was funded in whole or in part by the Austrian Science Fund (FWF) P 34671 and I 6068. We acknowledge the support of the Natural Sciences and Engineering Research Council of Canada (NSERC). 
\end{acknowledgments}


\bibliography{apssamp}

\begin{thebibliography}{37}%
\makeatletter
\providecommand \@ifxundefined [1]{%
 \@ifx{#1\undefined}
}%
\providecommand \@ifnum [1]{%
 \ifnum #1\expandafter \@firstoftwo
 \else \expandafter \@secondoftwo
 \fi
}%
\providecommand \@ifx [1]{%
 \ifx #1\expandafter \@firstoftwo
 \else \expandafter \@secondoftwo
 \fi
}%
\providecommand \natexlab [1]{#1}%
\providecommand \enquote  [1]{``#1''}%
\providecommand \bibnamefont  [1]{#1}%
\providecommand \bibfnamefont [1]{#1}%
\providecommand \citenamefont [1]{#1}%
\providecommand \href@noop [0]{\@secondoftwo}%
\providecommand \href [0]{\begingroup \@sanitize@url \@href}%
\providecommand \@href[1]{\@@startlink{#1}\@@href}%
\providecommand \@@href[1]{\endgroup#1\@@endlink}%
\providecommand \@sanitize@url [0]{\catcode `\\12\catcode `\$12\catcode `\&12\catcode `\#12\catcode `\^12\catcode `\_12\catcode `\%12\relax}%
\providecommand \@@startlink[1]{}%
\providecommand \@@endlink[0]{}%
\providecommand \url  [0]{\begingroup\@sanitize@url \@url }%
\providecommand \@url [1]{\endgroup\@href {#1}{\urlprefix }}%
\providecommand \urlprefix  [0]{URL }%
\providecommand \Eprint [0]{\href }%
\providecommand \doibase [0]{https://doi.org/}%
\providecommand \selectlanguage [0]{\@gobble}%
\providecommand \bibinfo  [0]{\@secondoftwo}%
\providecommand \bibfield  [0]{\@secondoftwo}%
\providecommand \translation [1]{[#1]}%
\providecommand \BibitemOpen [0]{}%
\providecommand \bibitemStop [0]{}%
\providecommand \bibitemNoStop [0]{.\EOS\space}%
\providecommand \EOS [0]{\spacefactor3000\relax}%
\providecommand \BibitemShut  [1]{\csname bibitem#1\endcsname}%
\let\auto@bib@innerbib\@empty
\bibitem [{\citenamefont {CHM}(2007)}]{chm_1970_2007}%
  \BibitemOpen
  \bibfield  {author} {\bibinfo {author} {\bibnamefont {CHM}},\ }\href {https://www.computerhistory.org/siliconengine/mos-dynamic-ram-competes-with-magnetic-core-memory-on-price/} {\bibinfo {title} {1970: {{MOS Dynamic RAM Competes}} with {{Magnetic Core Memory}} on {{Price}} {\textbar} {{The Silicon Engine}} {\textbar} {{Computer History Museum}}}} (\bibinfo {year} {2007})\BibitemShut {NoStop}%
\bibitem [{\citenamefont {Khvalkovskiy}\ \emph {et~al.}(2013)\citenamefont {Khvalkovskiy}, \citenamefont {Apalkov}, \citenamefont {Watts}, \citenamefont {Chepulskii}, \citenamefont {Beach}, \citenamefont {Ong}, \citenamefont {Tang}, \citenamefont {{Driskill-Smith}}, \citenamefont {Butler}, \citenamefont {Visscher}, \citenamefont {Lottis}, \citenamefont {Chen}, \citenamefont {Nikitin},\ and\ \citenamefont {Krounbi}}]{khvalkovskiy_basic_2013}%
  \BibitemOpen
  \bibfield  {author} {\bibinfo {author} {\bibfnamefont {A.~V.}\ \bibnamefont {Khvalkovskiy}}, \bibinfo {author} {\bibfnamefont {D.}~\bibnamefont {Apalkov}}, \bibinfo {author} {\bibfnamefont {S.}~\bibnamefont {Watts}}, \bibinfo {author} {\bibfnamefont {R.}~\bibnamefont {Chepulskii}}, \bibinfo {author} {\bibfnamefont {R.~S.}\ \bibnamefont {Beach}}, \bibinfo {author} {\bibfnamefont {A.}~\bibnamefont {Ong}}, \bibinfo {author} {\bibfnamefont {X.}~\bibnamefont {Tang}}, \bibinfo {author} {\bibfnamefont {A.}~\bibnamefont {{Driskill-Smith}}}, \bibinfo {author} {\bibfnamefont {W.~H.}\ \bibnamefont {Butler}}, \bibinfo {author} {\bibfnamefont {P.~B.}\ \bibnamefont {Visscher}}, \bibinfo {author} {\bibfnamefont {D.}~\bibnamefont {Lottis}}, \bibinfo {author} {\bibfnamefont {E.}~\bibnamefont {Chen}}, \bibinfo {author} {\bibfnamefont {V.}~\bibnamefont {Nikitin}},\ and\ \bibinfo {author} {\bibfnamefont {M.}~\bibnamefont {Krounbi}},\ }\bibfield  {title} {\bibinfo {title} {Basic principles of {{STT-MRAM}} cell operation in
  memory arrays},\ }\href {https://doi.org/10.1088/0022-3727/46/7/074001} {\bibfield  {journal} {\bibinfo  {journal} {Journal of Physics D: Applied Physics}\ }\textbf {\bibinfo {volume} {46}},\ \bibinfo {pages} {074001} (\bibinfo {year} {2013})}\BibitemShut {NoStop}%
\bibitem [{\citenamefont {Mangin}\ \emph {et~al.}(2006)\citenamefont {Mangin}, \citenamefont {Ravelosona}, \citenamefont {Katine}, \citenamefont {Carey}, \citenamefont {Terris},\ and\ \citenamefont {Fullerton}}]{mangin_current-induced_2006}%
  \BibitemOpen
  \bibfield  {author} {\bibinfo {author} {\bibfnamefont {S.}~\bibnamefont {Mangin}}, \bibinfo {author} {\bibfnamefont {D.}~\bibnamefont {Ravelosona}}, \bibinfo {author} {\bibfnamefont {J.~A.}\ \bibnamefont {Katine}}, \bibinfo {author} {\bibfnamefont {M.~J.}\ \bibnamefont {Carey}}, \bibinfo {author} {\bibfnamefont {B.~D.}\ \bibnamefont {Terris}},\ and\ \bibinfo {author} {\bibfnamefont {E.~E.}\ \bibnamefont {Fullerton}},\ }\bibfield  {title} {\bibinfo {title} {Current-induced magnetization reversal in nanopillars with perpendicular anisotropy},\ }\href {https://doi.org/10.1038/nmat1595} {\bibfield  {journal} {\bibinfo  {journal} {Nature Materials}\ }\textbf {\bibinfo {volume} {5}},\ \bibinfo {pages} {210} (\bibinfo {year} {2006})}\BibitemShut {NoStop}%
\bibitem [{\citenamefont {Weller}\ and\ \citenamefont {Moser}(1999)}]{weller_thermal_1999}%
  \BibitemOpen
  \bibfield  {author} {\bibinfo {author} {\bibfnamefont {D.}~\bibnamefont {Weller}}\ and\ \bibinfo {author} {\bibfnamefont {A.}~\bibnamefont {Moser}},\ }\bibfield  {title} {\bibinfo {title} {Thermal effect limits in ultrahigh-density magnetic recording},\ }\href {https://doi.org/10.1109/20.809134} {\bibfield  {journal} {\bibinfo  {journal} {IEEE Transactions on Magnetics}\ }\textbf {\bibinfo {volume} {35}},\ \bibinfo {pages} {4423} (\bibinfo {year} {1999})}\BibitemShut {NoStop}%
\bibitem [{\citenamefont {Slonczewski}(1996)}]{slonczewski_current-driven_1996}%
  \BibitemOpen
  \bibfield  {author} {\bibinfo {author} {\bibfnamefont {J.~C.}\ \bibnamefont {Slonczewski}},\ }\bibfield  {title} {\bibinfo {title} {Current-driven excitation of magnetic multilayers},\ }\href {https://doi.org/10.1016/0304-8853(96)00062-5} {\bibfield  {journal} {\bibinfo  {journal} {Journal of Magnetism and Magnetic Materials}\ }\textbf {\bibinfo {volume} {159}},\ \bibinfo {pages} {L1} (\bibinfo {year} {1996})}\BibitemShut {NoStop}%
\bibitem [{\citenamefont {Richter}\ and\ \citenamefont {Girt}(2002)}]{richter_how_2002}%
  \BibitemOpen
  \bibfield  {author} {\bibinfo {author} {\bibfnamefont {H.}~\bibnamefont {Richter}}\ and\ \bibinfo {author} {\bibfnamefont {E.}~\bibnamefont {Girt}},\ }\bibfield  {title} {\bibinfo {title} {How antiferromagnetic coupling can stabilize recorded information},\ }\href {https://doi.org/10.1109/TMAG.2002.801785} {\bibfield  {journal} {\bibinfo  {journal} {IEEE Transactions on Magnetics}\ }\textbf {\bibinfo {volume} {38}},\ \bibinfo {pages} {1867} (\bibinfo {year} {2002})}\BibitemShut {NoStop}%
\bibitem [{\citenamefont {Richter}\ \emph {et~al.}(2002)\citenamefont {Richter}, \citenamefont {Girt},\ and\ \citenamefont {Zhou}}]{richter_simplified_2002}%
  \BibitemOpen
  \bibfield  {author} {\bibinfo {author} {\bibfnamefont {H.~J.}\ \bibnamefont {Richter}}, \bibinfo {author} {\bibfnamefont {{\relax Er}.}~\bibnamefont {Girt}},\ and\ \bibinfo {author} {\bibfnamefont {H.}~\bibnamefont {Zhou}},\ }\bibfield  {title} {\bibinfo {title} {Simplified analysis of two-layer antiferromagnetically coupled media},\ }\href {https://doi.org/10.1063/1.1467977} {\bibfield  {journal} {\bibinfo  {journal} {Applied Physics Letters}\ }\textbf {\bibinfo {volume} {80}},\ \bibinfo {pages} {2529} (\bibinfo {year} {2002})}\BibitemShut {NoStop}%
\bibitem [{\citenamefont {Everspin}(2008)}]{everspin_everspin_2008}%
  \BibitemOpen
  \bibfield  {author} {\bibinfo {author} {\bibnamefont {Everspin}},\ }\href {https://www.everspin.com/} {\bibinfo {title} {Everspin {\textbar} {{The MRAM Company}}}} (\bibinfo {year} {2008})\BibitemShut {NoStop}%
\bibitem [{\citenamefont {Chung}\ \emph {et~al.}(2016)\citenamefont {Chung}, \citenamefont {Kishi}, \citenamefont {Park}, \citenamefont {Yoshikawa}, \citenamefont {Park}, \citenamefont {Nagase}, \citenamefont {Sunouchi}, \citenamefont {Kanaya}, \citenamefont {Kim}, \citenamefont {Noma}, \citenamefont {Lee}, \citenamefont {Yamamoto}, \citenamefont {Rho}, \citenamefont {Tsuchida}, \citenamefont {Chung}, \citenamefont {Yi}, \citenamefont {Kim}, \citenamefont {Chun}, \citenamefont {Oyamatsu},\ and\ \citenamefont {Hong}}]{chung_4gbit_2016}%
  \BibitemOpen
  \bibfield  {author} {\bibinfo {author} {\bibfnamefont {S.-W.}\ \bibnamefont {Chung}}, \bibinfo {author} {\bibfnamefont {T.}~\bibnamefont {Kishi}}, \bibinfo {author} {\bibfnamefont {J.~W.}\ \bibnamefont {Park}}, \bibinfo {author} {\bibfnamefont {M.}~\bibnamefont {Yoshikawa}}, \bibinfo {author} {\bibfnamefont {K.~S.}\ \bibnamefont {Park}}, \bibinfo {author} {\bibfnamefont {T.}~\bibnamefont {Nagase}}, \bibinfo {author} {\bibfnamefont {K.}~\bibnamefont {Sunouchi}}, \bibinfo {author} {\bibfnamefont {H.}~\bibnamefont {Kanaya}}, \bibinfo {author} {\bibfnamefont {G.~C.}\ \bibnamefont {Kim}}, \bibinfo {author} {\bibfnamefont {K.}~\bibnamefont {Noma}}, \bibinfo {author} {\bibfnamefont {M.~S.}\ \bibnamefont {Lee}}, \bibinfo {author} {\bibfnamefont {A.}~\bibnamefont {Yamamoto}}, \bibinfo {author} {\bibfnamefont {K.~M.}\ \bibnamefont {Rho}}, \bibinfo {author} {\bibfnamefont {K.}~\bibnamefont {Tsuchida}}, \bibinfo {author} {\bibfnamefont {S.~J.}\ \bibnamefont {Chung}}, \bibinfo {author} {\bibfnamefont {J.~Y.}\
  \bibnamefont {Yi}}, \bibinfo {author} {\bibfnamefont {H.~S.}\ \bibnamefont {Kim}}, \bibinfo {author} {\bibfnamefont {Y.}~\bibnamefont {Chun}}, \bibinfo {author} {\bibfnamefont {H.}~\bibnamefont {Oyamatsu}},\ and\ \bibinfo {author} {\bibfnamefont {S.~J.}\ \bibnamefont {Hong}},\ }\bibfield  {title} {\bibinfo {title} {{{4Gbit}} density {{STT-MRAM}} using perpendicular {{MTJ}} realized with compact cell structure},\ }in\ \href {https://doi.org/10.1109/IEDM.2016.7838490} {\emph {\bibinfo {booktitle} {2016 {{IEEE International Electron Devices Meeting}} ({{IEDM}})}}}\ (\bibinfo {year} {2016})\ pp.\ \bibinfo {pages} {27.1.1--27.1.4}\BibitemShut {NoStop}%
\bibitem [{\citenamefont {Kultursay}\ \emph {et~al.}(2013)\citenamefont {Kultursay}, \citenamefont {Kandemir}, \citenamefont {Sivasubramaniam},\ and\ \citenamefont {Mutlu}}]{kultursay_evaluating_2013}%
  \BibitemOpen
  \bibfield  {author} {\bibinfo {author} {\bibfnamefont {E.}~\bibnamefont {Kultursay}}, \bibinfo {author} {\bibfnamefont {M.}~\bibnamefont {Kandemir}}, \bibinfo {author} {\bibfnamefont {A.}~\bibnamefont {Sivasubramaniam}},\ and\ \bibinfo {author} {\bibfnamefont {O.}~\bibnamefont {Mutlu}},\ }\bibfield  {title} {\bibinfo {title} {Evaluating {{STT-RAM}} as an energy-efficient main memory alternative},\ }in\ \href {https://doi.org/10.1109/ISPASS.2013.6557176} {\emph {\bibinfo {booktitle} {2013 {{IEEE International Symposium}} on {{Performance Analysis}} of {{Systems}} and {{Software}} ({{ISPASS}})}}}\ (\bibinfo  {publisher} {{IEEE}},\ \bibinfo {address} {{Austin, TX, USA}},\ \bibinfo {year} {2013})\ pp.\ \bibinfo {pages} {256--267}\BibitemShut {NoStop}%
\bibitem [{\citenamefont {Lee}\ \emph {et~al.}(2020)\citenamefont {Lee}, \citenamefont {Yamane}, \citenamefont {Kwon}, \citenamefont {Naik}, \citenamefont {Otani}, \citenamefont {Zeng}, \citenamefont {Lim}, \citenamefont {Sivabalan}, \citenamefont {Chiang}, \citenamefont {Huang}, \citenamefont {Jang}, \citenamefont {Hau}, \citenamefont {Chao}, \citenamefont {Chung}, \citenamefont {Neo}, \citenamefont {Khua}, \citenamefont {Thiyagarajah}, \citenamefont {Ling}, \citenamefont {Goh}, \citenamefont {Hwang}, \citenamefont {Zhang}, \citenamefont {Low}, \citenamefont {Balasankaran}, \citenamefont {Tan}, \citenamefont {Wong}, \citenamefont {Seet}, \citenamefont {Ting}, \citenamefont {Ong}, \citenamefont {You}, \citenamefont {Woo},\ and\ \citenamefont {Siah}}]{lee_fast_2020}%
  \BibitemOpen
  \bibfield  {author} {\bibinfo {author} {\bibfnamefont {T.~Y.}\ \bibnamefont {Lee}}, \bibinfo {author} {\bibfnamefont {K.}~\bibnamefont {Yamane}}, \bibinfo {author} {\bibfnamefont {J.}~\bibnamefont {Kwon}}, \bibinfo {author} {\bibfnamefont {V.~B.}\ \bibnamefont {Naik}}, \bibinfo {author} {\bibfnamefont {Y.}~\bibnamefont {Otani}}, \bibinfo {author} {\bibfnamefont {D.}~\bibnamefont {Zeng}}, \bibinfo {author} {\bibfnamefont {J.~H.}\ \bibnamefont {Lim}}, \bibinfo {author} {\bibfnamefont {K.}~\bibnamefont {Sivabalan}}, \bibinfo {author} {\bibfnamefont {C.}~\bibnamefont {Chiang}}, \bibinfo {author} {\bibfnamefont {Y.}~\bibnamefont {Huang}}, \bibinfo {author} {\bibfnamefont {S.~H.}\ \bibnamefont {Jang}}, \bibinfo {author} {\bibfnamefont {L.~Y.}\ \bibnamefont {Hau}}, \bibinfo {author} {\bibfnamefont {R.}~\bibnamefont {Chao}}, \bibinfo {author} {\bibfnamefont {N.~L.}\ \bibnamefont {Chung}}, \bibinfo {author} {\bibfnamefont {W.~P.}\ \bibnamefont {Neo}}, \bibinfo {author} {\bibfnamefont {K.}~\bibnamefont {Khua}},
  \bibinfo {author} {\bibfnamefont {N.}~\bibnamefont {Thiyagarajah}}, \bibinfo {author} {\bibfnamefont {T.}~\bibnamefont {Ling}}, \bibinfo {author} {\bibfnamefont {L.~C.}\ \bibnamefont {Goh}}, \bibinfo {author} {\bibfnamefont {J.}~\bibnamefont {Hwang}}, \bibinfo {author} {\bibfnamefont {L.}~\bibnamefont {Zhang}}, \bibinfo {author} {\bibfnamefont {R.}~\bibnamefont {Low}}, \bibinfo {author} {\bibfnamefont {N.}~\bibnamefont {Balasankaran}}, \bibinfo {author} {\bibfnamefont {F.}~\bibnamefont {Tan}}, \bibinfo {author} {\bibfnamefont {J.}~\bibnamefont {Wong}}, \bibinfo {author} {\bibfnamefont {C.~S.}\ \bibnamefont {Seet}}, \bibinfo {author} {\bibfnamefont {J.~W.}\ \bibnamefont {Ting}}, \bibinfo {author} {\bibfnamefont {S.}~\bibnamefont {Ong}}, \bibinfo {author} {\bibfnamefont {Y.~S.}\ \bibnamefont {You}}, \bibinfo {author} {\bibfnamefont {S.~T.}\ \bibnamefont {Woo}},\ and\ \bibinfo {author} {\bibfnamefont {S.~Y.}\ \bibnamefont {Siah}},\ }\bibfield  {title} {\bibinfo {title} {Fast {{Switching}} of {{STT-MRAM}} to
  {{Realize High Speed Applications}}},\ }in\ \href {https://doi.org/10.1109/VLSITechnology18217.2020.9265027} {\emph {\bibinfo {booktitle} {2020 {{IEEE Symposium}} on {{VLSI Technology}}}}}\ (\bibinfo  {publisher} {{IEEE}},\ \bibinfo {address} {{Honolulu, HI, USA}},\ \bibinfo {year} {2020})\ pp.\ \bibinfo {pages} {1--2}\BibitemShut {NoStop}%
\bibitem [{\citenamefont {Safranski}\ \emph {et~al.}(2022)\citenamefont {Safranski}, \citenamefont {Hu}, \citenamefont {Sun}, \citenamefont {Hashemi}, \citenamefont {Brown}, \citenamefont {Buzi}, \citenamefont {D'Emic}, \citenamefont {Edwards}, \citenamefont {Galligan}, \citenamefont {Gottwald}, \citenamefont {Gunawan}, \citenamefont {Karimeddiny}, \citenamefont {Jung}, \citenamefont {Kim}, \citenamefont {Latzko}, \citenamefont {Trouilloud},\ and\ \citenamefont {Worledge}}]{safranski_reliable_2022}%
  \BibitemOpen
  \bibfield  {author} {\bibinfo {author} {\bibfnamefont {C.}~\bibnamefont {Safranski}}, \bibinfo {author} {\bibfnamefont {G.}~\bibnamefont {Hu}}, \bibinfo {author} {\bibfnamefont {J.~Z.}\ \bibnamefont {Sun}}, \bibinfo {author} {\bibfnamefont {P.}~\bibnamefont {Hashemi}}, \bibinfo {author} {\bibfnamefont {S.~L.}\ \bibnamefont {Brown}}, \bibinfo {author} {\bibfnamefont {L.}~\bibnamefont {Buzi}}, \bibinfo {author} {\bibfnamefont {C.~P.}\ \bibnamefont {D'Emic}}, \bibinfo {author} {\bibfnamefont {E.~R.~J.}\ \bibnamefont {Edwards}}, \bibinfo {author} {\bibfnamefont {E.}~\bibnamefont {Galligan}}, \bibinfo {author} {\bibfnamefont {M.~G.}\ \bibnamefont {Gottwald}}, \bibinfo {author} {\bibfnamefont {O.}~\bibnamefont {Gunawan}}, \bibinfo {author} {\bibfnamefont {S.}~\bibnamefont {Karimeddiny}}, \bibinfo {author} {\bibfnamefont {H.}~\bibnamefont {Jung}}, \bibinfo {author} {\bibfnamefont {J.}~\bibnamefont {Kim}}, \bibinfo {author} {\bibfnamefont {K.}~\bibnamefont {Latzko}}, \bibinfo {author} {\bibfnamefont {P.~L.}\
  \bibnamefont {Trouilloud}},\ and\ \bibinfo {author} {\bibfnamefont {D.~C.}\ \bibnamefont {Worledge}},\ }\bibfield  {title} {\bibinfo {title} {Reliable {{Sub-Nanosecond Switching}} in {{Magnetic Tunnel Junctions}} for {{MRAM Applications}}},\ }\href {https://doi.org/10.1109/TED.2022.3214168} {\bibfield  {journal} {\bibinfo  {journal} {IEEE Transactions on Electron Devices}\ }\textbf {\bibinfo {volume} {69}},\ \bibinfo {pages} {7180} (\bibinfo {year} {2022})}\BibitemShut {NoStop}%
\bibitem [{\citenamefont {Kent}\ and\ \citenamefont {Worledge}(2015)}]{kent_new_2015}%
  \BibitemOpen
  \bibfield  {author} {\bibinfo {author} {\bibfnamefont {A.~D.}\ \bibnamefont {Kent}}\ and\ \bibinfo {author} {\bibfnamefont {D.~C.}\ \bibnamefont {Worledge}},\ }\bibfield  {title} {\bibinfo {title} {A new spin on magnetic memories},\ }\href {https://doi.org/10.1038/nnano.2015.24} {\bibfield  {journal} {\bibinfo  {journal} {Nature Nanotechnology}\ }\textbf {\bibinfo {volume} {10}},\ \bibinfo {pages} {187} (\bibinfo {year} {2015})}\BibitemShut {NoStop}%
\bibitem [{\citenamefont {Worledge}(2023)}]{worledge_write-error-rate_2023}%
  \BibitemOpen
  \bibfield  {author} {\bibinfo {author} {\bibfnamefont {D.~C.}\ \bibnamefont {Worledge}},\ }\bibfield  {title} {\bibinfo {title} {Write-error-rate of {{Spin-Transfer-Torque MRAM}} ({{Invited}})},\ }in\ \href {https://doi.org/10.1109/IRPS48203.2023.10117666} {\emph {\bibinfo {booktitle} {2023 {{IEEE International Reliability Physics Symposium}} ({{IRPS}})}}}\ (\bibinfo {year} {2023})\ pp.\ \bibinfo {pages} {1--4}\BibitemShut {NoStop}%
\bibitem [{\citenamefont {Sbiaa}(2013)}]{sbiaa_magnetization_2013}%
  \BibitemOpen
  \bibfield  {author} {\bibinfo {author} {\bibfnamefont {R.}~\bibnamefont {Sbiaa}},\ }\bibfield  {title} {\bibinfo {title} {Magnetization switching by spin-torque effect in off-aligned structure with perpendicular anisotropy},\ }\href {https://doi.org/10.1088/0022-3727/46/39/395001} {\bibfield  {journal} {\bibinfo  {journal} {Journal of Physics D: Applied Physics}\ }\textbf {\bibinfo {volume} {46}},\ \bibinfo {pages} {395001} (\bibinfo {year} {2013})}\BibitemShut {NoStop}%
\bibitem [{\citenamefont {Matsumoto}\ \emph {et~al.}(2015)\citenamefont {Matsumoto}, \citenamefont {Arai}, \citenamefont {Yuasa},\ and\ \citenamefont {Imamura}}]{matsumoto_spin-transfer-torque_2015}%
  \BibitemOpen
  \bibfield  {author} {\bibinfo {author} {\bibfnamefont {R.}~\bibnamefont {Matsumoto}}, \bibinfo {author} {\bibfnamefont {H.}~\bibnamefont {Arai}}, \bibinfo {author} {\bibfnamefont {S.}~\bibnamefont {Yuasa}},\ and\ \bibinfo {author} {\bibfnamefont {H.}~\bibnamefont {Imamura}},\ }\bibfield  {title} {\bibinfo {title} {Spin-transfer-torque switching in a spin-valve nanopillar with a conically magnetized free layer},\ }\href {https://doi.org/10.7567/APEX.8.063007} {\bibfield  {journal} {\bibinfo  {journal} {Applied Physics Express}\ }\textbf {\bibinfo {volume} {8}},\ \bibinfo {pages} {063007} (\bibinfo {year} {2015})}\BibitemShut {NoStop}%
\bibitem [{\citenamefont {Nunn}\ \emph {et~al.}(2020)\citenamefont {Nunn}, \citenamefont {Abert}, \citenamefont {Suess},\ and\ \citenamefont {Girt}}]{nunn_control_2020}%
  \BibitemOpen
  \bibfield  {author} {\bibinfo {author} {\bibfnamefont {Z.~R.}\ \bibnamefont {Nunn}}, \bibinfo {author} {\bibfnamefont {C.}~\bibnamefont {Abert}}, \bibinfo {author} {\bibfnamefont {D.}~\bibnamefont {Suess}},\ and\ \bibinfo {author} {\bibfnamefont {E.}~\bibnamefont {Girt}},\ }\bibfield  {title} {\bibinfo {title} {Control of the noncollinear interlayer exchange coupling},\ }\href {https://doi.org/10.1126/sciadv.abd8861} {\bibfield  {journal} {\bibinfo  {journal} {Science Advances}\ }\textbf {\bibinfo {volume} {6}},\ \bibinfo {pages} {eabd8861} (\bibinfo {year} {2020})}\BibitemShut {NoStop}%
\bibitem [{\citenamefont {Stiles}(2005)}]{stiles_interlayer_2005}%
  \BibitemOpen
  \bibfield  {author} {\bibinfo {author} {\bibfnamefont {M.}~\bibnamefont {Stiles}},\ }\bibfield  {title} {\bibinfo {title} {Interlayer {{Exchange Coupling}}},\ }in\ \href {https://doi.org/10.1007/3-540-27163-5_4} {\emph {\bibinfo {booktitle} {Ultrathin {{Magnetic Structures III}}: {{Fundamentals}} of {{Nanomagnetism}}}}},\ \bibinfo {editor} {edited by\ \bibinfo {editor} {\bibfnamefont {J.~A.~C.}\ \bibnamefont {Bland}}\ and\ \bibinfo {editor} {\bibfnamefont {B.}~\bibnamefont {Heinrich}}}\ (\bibinfo  {publisher} {{Springer}},\ \bibinfo {address} {{Berlin, Heidelberg}},\ \bibinfo {year} {2005})\ pp.\ \bibinfo {pages} {99--142}\BibitemShut {NoStop}%
\bibitem [{\citenamefont {Slonczewski}(1991)}]{slonczewski_fluctuation_1991}%
  \BibitemOpen
  \bibfield  {author} {\bibinfo {author} {\bibfnamefont {J.~C.}\ \bibnamefont {Slonczewski}},\ }\bibfield  {title} {\bibinfo {title} {Fluctuation mechanism for biquadratic exchange coupling in magnetic multilayers},\ }\href {https://doi.org/10.1103/PhysRevLett.67.3172} {\bibfield  {journal} {\bibinfo  {journal} {Physical Review Letters}\ }\textbf {\bibinfo {volume} {67}},\ \bibinfo {pages} {3172} (\bibinfo {year} {1991})}\BibitemShut {NoStop}%
\bibitem [{\citenamefont {Nunn}\ and\ \citenamefont {Girt}(2019)}]{nunn_non-collinear_2019}%
  \BibitemOpen
  \bibfield  {author} {\bibinfo {author} {\bibfnamefont {Z.~R.}\ \bibnamefont {Nunn}}\ and\ \bibinfo {author} {\bibfnamefont {E.}~\bibnamefont {Girt}},\ }\href {https://doi.org/10.48550/arXiv.1901.07055} {\bibinfo {title} {Non-collinear coupling across {{RuCo}} and {{RuFe}} alloys}} (\bibinfo {year} {2019}),\ \Eprint {https://arxiv.org/abs/1901.07055} {arxiv:1901.07055 [cond-mat]} \BibitemShut {NoStop}%
\bibitem [{\citenamefont {Bland}\ and\ \citenamefont {Heinrich}(2005)}]{bland_ultrathin_2005}%
  \BibitemOpen
  \bibfield  {author} {\bibinfo {author} {\bibfnamefont {J.~A.~C.}\ \bibnamefont {Bland}}\ and\ \bibinfo {author} {\bibfnamefont {B.}~\bibnamefont {Heinrich}},\ }\href {https://doi.org/10.1007/b138703} {\emph {\bibinfo {title} {Ultrathin {{Magnetic Structures III Fundamentals}} of {{Nanomagnetism}}}}},\ \bibinfo {edition} {1st}\ ed.,\ \bibinfo {series} {Ultrathin Magnetic Structures}\ No.~\bibinfo {number} {3}\ (\bibinfo  {publisher} {{Springer Berlin Heidelberg}},\ \bibinfo {address} {{Berlin, Heidelberg}},\ \bibinfo {year} {2005})\BibitemShut {NoStop}%
\bibitem [{\citenamefont {Bobo}\ \emph {et~al.}(1999)\citenamefont {Bobo}, \citenamefont {Kikuchi}, \citenamefont {Redon}, \citenamefont {Snoeck}, \citenamefont {Piecuch},\ and\ \citenamefont {White}}]{bobo_pinholes_1999}%
  \BibitemOpen
  \bibfield  {author} {\bibinfo {author} {\bibfnamefont {J.~F.}\ \bibnamefont {Bobo}}, \bibinfo {author} {\bibfnamefont {H.}~\bibnamefont {Kikuchi}}, \bibinfo {author} {\bibfnamefont {O.}~\bibnamefont {Redon}}, \bibinfo {author} {\bibfnamefont {E.}~\bibnamefont {Snoeck}}, \bibinfo {author} {\bibfnamefont {M.}~\bibnamefont {Piecuch}},\ and\ \bibinfo {author} {\bibfnamefont {R.~L.}\ \bibnamefont {White}},\ }\bibfield  {title} {\bibinfo {title} {Pinholes in antiferromagnetically coupled multilayers: {{Effects}} on hysteresis loops and relation to biquadratic exchange},\ }\href {https://doi.org/10.1103/PhysRevB.60.4131} {\bibfield  {journal} {\bibinfo  {journal} {Physical Review B}\ }\textbf {\bibinfo {volume} {60}},\ \bibinfo {pages} {4131} (\bibinfo {year} {1999})}\BibitemShut {NoStop}%
\bibitem [{\citenamefont {Slonczewski}(1993)}]{slonczewski_origin_1993}%
  \BibitemOpen
  \bibfield  {author} {\bibinfo {author} {\bibfnamefont {J.~C.}\ \bibnamefont {Slonczewski}},\ }\bibfield  {title} {\bibinfo {title} {Origin of biquadratic exchange in magnetic multilayers (invited)},\ }\href {https://doi.org/10.1063/1.353483} {\bibfield  {journal} {\bibinfo  {journal} {Journal of Applied Physics}\ }\textbf {\bibinfo {volume} {73}},\ \bibinfo {pages} {5957} (\bibinfo {year} {1993})}\BibitemShut {NoStop}%
\bibitem [{\citenamefont {Nunn}\ \emph {et~al.}(2023)\citenamefont {Nunn}, \citenamefont {Besler}, \citenamefont {Omelchenko}, \citenamefont {Koraltan}, \citenamefont {Abert}, \citenamefont {Suess},\ and\ \citenamefont {Girt}}]{nunn_controlling_2023}%
  \BibitemOpen
  \bibfield  {author} {\bibinfo {author} {\bibfnamefont {Z.~R.}\ \bibnamefont {Nunn}}, \bibinfo {author} {\bibfnamefont {J.}~\bibnamefont {Besler}}, \bibinfo {author} {\bibfnamefont {P.}~\bibnamefont {Omelchenko}}, \bibinfo {author} {\bibfnamefont {S.}~\bibnamefont {Koraltan}}, \bibinfo {author} {\bibfnamefont {C.}~\bibnamefont {Abert}}, \bibinfo {author} {\bibfnamefont {D.}~\bibnamefont {Suess}},\ and\ \bibinfo {author} {\bibfnamefont {E.}~\bibnamefont {Girt}},\ }\bibfield  {title} {\bibinfo {title} {Controlling the angle between magnetic moments of {{Co}} layers in {{Co}}{\textbar}{{RuCo}}{\textbar}{{Co}}},\ }\href {https://doi.org/10.1063/5.0141180} {\bibfield  {journal} {\bibinfo  {journal} {Journal of Applied Physics}\ }\textbf {\bibinfo {volume} {133}},\ \bibinfo {pages} {123901} (\bibinfo {year} {2023})}\BibitemShut {NoStop}%
\bibitem [{\citenamefont {Besler}\ \emph {et~al.}(2023)\citenamefont {Besler}, \citenamefont {Myrtle},\ and\ \citenamefont {Girt}}]{besler_noncollinear_2023}%
  \BibitemOpen
  \bibfield  {author} {\bibinfo {author} {\bibfnamefont {J.}~\bibnamefont {Besler}}, \bibinfo {author} {\bibfnamefont {S.}~\bibnamefont {Myrtle}},\ and\ \bibinfo {author} {\bibfnamefont {E.}~\bibnamefont {Girt}},\ }\href {https://doi.org/10.2139/ssrn.4442160} {\bibinfo {title} {Noncollinear {{Interlayer Exchange Coupling Across Irfe Spacer Layers}}}} (\bibinfo {year} {2023})\BibitemShut {NoStop}%
\bibitem [{\citenamefont {Abert}\ \emph {et~al.}(2022)\citenamefont {Abert}, \citenamefont {Koraltan}, \citenamefont {Bruckner}, \citenamefont {Slanovc}, \citenamefont {Lisik}, \citenamefont {Omelchenko}, \citenamefont {Girt},\ and\ \citenamefont {Suess}}]{abert_origin_2022}%
  \BibitemOpen
  \bibfield  {author} {\bibinfo {author} {\bibfnamefont {C.}~\bibnamefont {Abert}}, \bibinfo {author} {\bibfnamefont {S.}~\bibnamefont {Koraltan}}, \bibinfo {author} {\bibfnamefont {F.}~\bibnamefont {Bruckner}}, \bibinfo {author} {\bibfnamefont {F.}~\bibnamefont {Slanovc}}, \bibinfo {author} {\bibfnamefont {J.}~\bibnamefont {Lisik}}, \bibinfo {author} {\bibfnamefont {P.}~\bibnamefont {Omelchenko}}, \bibinfo {author} {\bibfnamefont {E.}~\bibnamefont {Girt}},\ and\ \bibinfo {author} {\bibfnamefont {D.}~\bibnamefont {Suess}},\ }\bibfield  {title} {\bibinfo {title} {Origin of noncollinear magnetization coupling across {{Ru X}} layers},\ }\href {https://doi.org/10.1103/PhysRevB.106.054401} {\bibfield  {journal} {\bibinfo  {journal} {Physical Review B}\ }\textbf {\bibinfo {volume} {106}},\ \bibinfo {pages} {054401} (\bibinfo {year} {2022})}\BibitemShut {NoStop}%
\bibitem [{\citenamefont {Arora}(2017)}]{arora_origin_2017}%
  \BibitemOpen
  \bibfield  {author} {\bibinfo {author} {\bibfnamefont {M.}~\bibnamefont {Arora}},\ }\emph {\bibinfo {title} {Origin of Perpendicular Magnetic Anisotropy in {{Co}}/{{Ni}} Multilayers and Their Use in {{STT-RAM}}}},\ \href {https://summit.sfu.ca/item/17855} {Ph.D. thesis},\ \bibinfo  {school} {Simon Fraser University} (\bibinfo {year} {2017})\BibitemShut {NoStop}%
\bibitem [{\citenamefont {Abert}\ \emph {et~al.}(2013)\citenamefont {Abert}, \citenamefont {Exl}, \citenamefont {Bruckner}, \citenamefont {Drews},\ and\ \citenamefont {Suess}}]{abert_magnumfe_2013}%
  \BibitemOpen
  \bibfield  {author} {\bibinfo {author} {\bibfnamefont {C.}~\bibnamefont {Abert}}, \bibinfo {author} {\bibfnamefont {L.}~\bibnamefont {Exl}}, \bibinfo {author} {\bibfnamefont {F.}~\bibnamefont {Bruckner}}, \bibinfo {author} {\bibfnamefont {A.}~\bibnamefont {Drews}},\ and\ \bibinfo {author} {\bibfnamefont {D.}~\bibnamefont {Suess}},\ }\bibfield  {title} {\bibinfo {title} {Magnum.fe: {{A}} micromagnetic finite-element simulation code based on {{FEniCS}}},\ }\href {https://doi.org/10.1016/j.jmmm.2013.05.051} {\bibfield  {journal} {\bibinfo  {journal} {Journal of Magnetism and Magnetic Materials}\ }\textbf {\bibinfo {volume} {345}},\ \bibinfo {pages} {29} (\bibinfo {year} {2013})}\BibitemShut {NoStop}%
\bibitem [{\citenamefont {Abert}(2019)}]{abert_micromagnetics_2019}%
  \BibitemOpen
  \bibfield  {author} {\bibinfo {author} {\bibfnamefont {C.}~\bibnamefont {Abert}},\ }\bibfield  {title} {\bibinfo {title} {Micromagnetics and spintronics: Models and numerical methods},\ }\href {https://doi.org/10.1140/epjb/e2019-90599-6} {\bibfield  {journal} {\bibinfo  {journal} {The European Physical Journal B}\ }\textbf {\bibinfo {volume} {92}},\ \bibinfo {pages} {120} (\bibinfo {year} {2019})}\BibitemShut {NoStop}%
\bibitem [{\citenamefont {Suess}\ \emph {et~al.}(2023)\citenamefont {Suess}, \citenamefont {Koraltan}, \citenamefont {Slanovc}, \citenamefont {Bruckner},\ and\ \citenamefont {Abert}}]{suess_accurate_2023}%
  \BibitemOpen
  \bibfield  {author} {\bibinfo {author} {\bibfnamefont {D.}~\bibnamefont {Suess}}, \bibinfo {author} {\bibfnamefont {S.}~\bibnamefont {Koraltan}}, \bibinfo {author} {\bibfnamefont {F.}~\bibnamefont {Slanovc}}, \bibinfo {author} {\bibfnamefont {F.}~\bibnamefont {Bruckner}},\ and\ \bibinfo {author} {\bibfnamefont {C.}~\bibnamefont {Abert}},\ }\bibfield  {title} {\bibinfo {title} {Accurate finite-difference micromagnetics of magnets including {{RKKY}} interaction: {{Analytical}} solution and comparison to standard micromagnetic codes},\ }\href {https://doi.org/10.1103/PhysRevB.107.104424} {\bibfield  {journal} {\bibinfo  {journal} {Physical Review B}\ }\textbf {\bibinfo {volume} {107}},\ \bibinfo {pages} {104424} (\bibinfo {year} {2023})}\BibitemShut {NoStop}%
\bibitem [{\citenamefont {{Lertzman-Lepofsky}}(2023)}]{lertzman-lepofsky_applications_2023}%
  \BibitemOpen
  \bibfield  {author} {\bibinfo {author} {\bibfnamefont {G.}~\bibnamefont {{Lertzman-Lepofsky}}},\ }\emph {\bibinfo {title} {Applications of {{Micromagnetic Simulations}} for {{Use}} in the {{Development}} of {{Novel Magnetic Computer Memory}}}},\ \href {https://summit.sfu.ca/item/35998} {\bibinfo {type} {Bachelor's thesis}},\ \bibinfo  {school} {Simon Fraser University} (\bibinfo {year} {2023})\BibitemShut {NoStop}%
\bibitem [{\citenamefont {Shimatsu}\ \emph {et~al.}(2004)\citenamefont {Shimatsu}, \citenamefont {Sato}, \citenamefont {Oikawa}, \citenamefont {Inaba}, \citenamefont {Kitakami}, \citenamefont {Okamoto}, \citenamefont {Aoi}, \citenamefont {Muraoka},\ and\ \citenamefont {Nakamura}}]{shimatsu_high_2004}%
  \BibitemOpen
  \bibfield  {author} {\bibinfo {author} {\bibfnamefont {T.}~\bibnamefont {Shimatsu}}, \bibinfo {author} {\bibfnamefont {H.}~\bibnamefont {Sato}}, \bibinfo {author} {\bibfnamefont {T.}~\bibnamefont {Oikawa}}, \bibinfo {author} {\bibfnamefont {Y.}~\bibnamefont {Inaba}}, \bibinfo {author} {\bibfnamefont {O.}~\bibnamefont {Kitakami}}, \bibinfo {author} {\bibfnamefont {S.}~\bibnamefont {Okamoto}}, \bibinfo {author} {\bibfnamefont {H.}~\bibnamefont {Aoi}}, \bibinfo {author} {\bibfnamefont {H.}~\bibnamefont {Muraoka}},\ and\ \bibinfo {author} {\bibfnamefont {Y.}~\bibnamefont {Nakamura}},\ }\bibfield  {title} {\bibinfo {title} {High {{Perpendicular Magnetic Anisotropy}} of {{CoPtCr}}/{{Ru Films}} for {{Granular-Type Perpendicular Media}}},\ }\href {https://doi.org/10.1109/TMAG.2004.832448} {\bibfield  {journal} {\bibinfo  {journal} {IEEE Transactions on Magnetics}\ }\textbf {\bibinfo {volume} {40}},\ \bibinfo {pages} {2483} (\bibinfo {year} {2004})}\BibitemShut {NoStop}%
\bibitem [{\citenamefont {Eyrich}\ \emph {et~al.}(2014)\citenamefont {Eyrich}, \citenamefont {Zamani}, \citenamefont {Huttema}, \citenamefont {Arora}, \citenamefont {Harrison}, \citenamefont {Rashidi}, \citenamefont {Broun}, \citenamefont {Heinrich}, \citenamefont {Mryasov}, \citenamefont {Ahlberg}, \citenamefont {Karis}, \citenamefont {J{\"o}nsson}, \citenamefont {From}, \citenamefont {Zhu},\ and\ \citenamefont {Girt}}]{eyrich_effects_2014}%
  \BibitemOpen
  \bibfield  {author} {\bibinfo {author} {\bibfnamefont {C.}~\bibnamefont {Eyrich}}, \bibinfo {author} {\bibfnamefont {A.}~\bibnamefont {Zamani}}, \bibinfo {author} {\bibfnamefont {W.}~\bibnamefont {Huttema}}, \bibinfo {author} {\bibfnamefont {M.}~\bibnamefont {Arora}}, \bibinfo {author} {\bibfnamefont {D.}~\bibnamefont {Harrison}}, \bibinfo {author} {\bibfnamefont {F.}~\bibnamefont {Rashidi}}, \bibinfo {author} {\bibfnamefont {D.}~\bibnamefont {Broun}}, \bibinfo {author} {\bibfnamefont {B.}~\bibnamefont {Heinrich}}, \bibinfo {author} {\bibfnamefont {O.}~\bibnamefont {Mryasov}}, \bibinfo {author} {\bibfnamefont {M.}~\bibnamefont {Ahlberg}}, \bibinfo {author} {\bibfnamefont {O.}~\bibnamefont {Karis}}, \bibinfo {author} {\bibfnamefont {P.~E.}\ \bibnamefont {J{\"o}nsson}}, \bibinfo {author} {\bibfnamefont {M.}~\bibnamefont {From}}, \bibinfo {author} {\bibfnamefont {X.}~\bibnamefont {Zhu}},\ and\ \bibinfo {author} {\bibfnamefont {E.}~\bibnamefont {Girt}},\ }\bibfield  {title} {\bibinfo {title} {Effects of
  substitution on the exchange stiffness and magnetization of {{Co}} films},\ }\href {https://doi.org/10.1103/PhysRevB.90.235408} {\bibfield  {journal} {\bibinfo  {journal} {Physical Review B}\ }\textbf {\bibinfo {volume} {90}},\ \bibinfo {pages} {235408} (\bibinfo {year} {2014})}\BibitemShut {NoStop}%
\bibitem [{\citenamefont {E}\ \emph {et~al.}(2007)\citenamefont {E}, \citenamefont {Ren},\ and\ \citenamefont {{Vanden-Eijnden}}}]{e_simplified_2007}%
  \BibitemOpen
  \bibfield  {author} {\bibinfo {author} {\bibfnamefont {W.}~\bibnamefont {E}}, \bibinfo {author} {\bibfnamefont {W.}~\bibnamefont {Ren}},\ and\ \bibinfo {author} {\bibfnamefont {E.}~\bibnamefont {{Vanden-Eijnden}}},\ }\bibfield  {title} {\bibinfo {title} {Simplified and improved string method for computing the minimum energy paths in barrier-crossing events},\ }\href {https://doi.org/10.1063/1.2720838} {\bibfield  {journal} {\bibinfo  {journal} {The Journal of Chemical Physics}\ }\textbf {\bibinfo {volume} {126}},\ \bibinfo {pages} {164103} (\bibinfo {year} {2007})}\BibitemShut {NoStop}%
\bibitem [{\citenamefont {Koraltan}\ \emph {et~al.}(2020)\citenamefont {Koraltan}, \citenamefont {Pancaldi}, \citenamefont {Leo}, \citenamefont {Abert}, \citenamefont {Vogler}, \citenamefont {Hofhuis}, \citenamefont {Slanovc}, \citenamefont {Bruckner}, \citenamefont {Heistracher}, \citenamefont {Menniti}, \citenamefont {Vavassori},\ and\ \citenamefont {Suess}}]{koraltan_dependence_2020}%
  \BibitemOpen
  \bibfield  {author} {\bibinfo {author} {\bibfnamefont {S.}~\bibnamefont {Koraltan}}, \bibinfo {author} {\bibfnamefont {M.}~\bibnamefont {Pancaldi}}, \bibinfo {author} {\bibfnamefont {N.}~\bibnamefont {Leo}}, \bibinfo {author} {\bibfnamefont {C.}~\bibnamefont {Abert}}, \bibinfo {author} {\bibfnamefont {C.}~\bibnamefont {Vogler}}, \bibinfo {author} {\bibfnamefont {K.}~\bibnamefont {Hofhuis}}, \bibinfo {author} {\bibfnamefont {F.}~\bibnamefont {Slanovc}}, \bibinfo {author} {\bibfnamefont {F.}~\bibnamefont {Bruckner}}, \bibinfo {author} {\bibfnamefont {P.}~\bibnamefont {Heistracher}}, \bibinfo {author} {\bibfnamefont {M.}~\bibnamefont {Menniti}}, \bibinfo {author} {\bibfnamefont {P.}~\bibnamefont {Vavassori}},\ and\ \bibinfo {author} {\bibfnamefont {D.}~\bibnamefont {Suess}},\ }\bibfield  {title} {\bibinfo {title} {Dependence of energy barrier reduction on collective excitations in square artificial spin ice: {{A}} comprehensive comparison of simulation techniques},\ }\href
  {https://doi.org/10.1103/PhysRevB.102.064410} {\bibfield  {journal} {\bibinfo  {journal} {Physical Review B}\ }\textbf {\bibinfo {volume} {102}},\ \bibinfo {pages} {064410} (\bibinfo {year} {2020})}\BibitemShut {NoStop}%
\bibitem [{\citenamefont {Grollier}\ \emph {et~al.}(2020)\citenamefont {Grollier}, \citenamefont {Querlioz}, \citenamefont {{\c C}amsar{\i}}, \citenamefont {{Everschor-Sitte}}, \citenamefont {Fukami},\ and\ \citenamefont {Stiles}}]{grollier_neuromorphic_2020}%
  \BibitemOpen
  \bibfield  {author} {\bibinfo {author} {\bibfnamefont {J.}~\bibnamefont {Grollier}}, \bibinfo {author} {\bibfnamefont {D.}~\bibnamefont {Querlioz}}, \bibinfo {author} {\bibfnamefont {K.}~\bibnamefont {{\c C}amsar{\i}}}, \bibinfo {author} {\bibfnamefont {K.}~\bibnamefont {{Everschor-Sitte}}}, \bibinfo {author} {\bibfnamefont {S.}~\bibnamefont {Fukami}},\ and\ \bibinfo {author} {\bibfnamefont {M.}~\bibnamefont {Stiles}},\ }\bibfield  {title} {\bibinfo {title} {Neuromorphic {{Spintronics}}},\ }\href {https://doi.org/10.1038/s41928-019-0360-9} {\bibfield  {journal} {\bibinfo  {journal} {Nature Electronics}\ }\textbf {\bibinfo {volume} {3}},\ \bibinfo {pages} {1} (\bibinfo {year} {2020})}\BibitemShut {NoStop}%
\bibitem [{\citenamefont {McKinnon}\ \emph {et~al.}(2022)\citenamefont {McKinnon}, \citenamefont {Heinrich},\ and\ \citenamefont {Girt}}]{mckinnon_thermally_2022}%
  \BibitemOpen
  \bibfield  {author} {\bibinfo {author} {\bibfnamefont {T.}~\bibnamefont {McKinnon}}, \bibinfo {author} {\bibfnamefont {B.}~\bibnamefont {Heinrich}},\ and\ \bibinfo {author} {\bibfnamefont {E.}~\bibnamefont {Girt}},\ }\bibfield  {title} {\bibinfo {title} {Thermally robust synthetic antiferromagnetic fixed layers containing {{FeCoB}} for use in {{STT-MRAM}} devices},\ }\href {https://doi.org/10.1016/j.jmmm.2021.168646} {\bibfield  {journal} {\bibinfo  {journal} {Journal of Magnetism and Magnetic Materials}\ }\textbf {\bibinfo {volume} {546}},\ \bibinfo {pages} {168646} (\bibinfo {year} {2022})}\BibitemShut {NoStop}%
\end{thebibliography}%

\end{document}